\documentclass[fleqn,usenatbib]{mnras}

\pdfoutput=1  

\usepackage{graphicx}    
\usepackage{amsmath}     
\usepackage{amssymb}     
\usepackage{gensymb}     
\usepackage{bm}          
\usepackage{xspace}      
\usepackage{nicefrac}    
\usepackage{hyperref}    
\usepackage{xcolor}      
\usepackage{ifthen}
\usepackage{enumitem}

\title[Lessons from a blind study of simulated lenses]{Lessons from a blind study of simulated lenses: \\ image reconstructions do not always reproduce true convergence}

\newcommand{\lens}[1]{\ifthenelse{\equal{#1}{0}}{H1S0A0B90G0}{}\ifthenelse{\equal{#1}{1}}{H1S1A0B90G0}{}\ifthenelse{\equal{#1}{2}}{H2S1A0B90G0}{}\ifthenelse{\equal{#1}{3}}{H2S2A0B90G0}{}\ifthenelse{\equal{#1}{4}}{H2S7A0B90G0}{}\ifthenelse{\equal{#1}{5}}{H3S0A0B90G0}{}\ifthenelse{\equal{#1}{6}}{H3S1A0B90G0}{}\ifthenelse{\equal{#1}{7}}{H4S3A0B0G90}{}\ifthenelse{\equal{#1}{8}}{H10S0A0B90G0}{}\ifthenelse{\equal{#1}{9}}{H13S0A0B90G0}{}\ifthenelse{\equal{#1}{10}}{H23S0A0B90G0}{}\ifthenelse{\equal{#1}{11}}{H30S0A0B90G0}{}\ifthenelse{\equal{#1}{12}}{H36S0A0B90G0}{}\ifthenelse{\equal{#1}{13}}{H160S0A90B0G0}{}\ifthenelse{\equal{#1}{14}}{H234S0A0B90G0}{}\ignorespaces}

\newcommand{\nauthor}[2]{#2,$^{#1}$}
\newcommand{\nauthorl}[2]{#2$^{#1}$}
\newcommand{\email}[1]{\thanks{Email: #1}}
\newcommand{\naffiliation}[2]{$^{#1}$#2}

\author[Denzel et al.]{%
  \nauthor{1,2}{Philipp Denzel} \email{phdenzel@physik.uzh.ch}
  \nauthor{3}{Sampath Mukherjee} 
  \nauthor{4}{Jonathan P. Coles} 
  \nauthorl{1,2}{Prasenjit Saha} 
  \newauthor
  \\
  \naffiliation{1}{Institute for Computational Science, University of Zurich, 8057 Zurich, Switzerland} \\
  \naffiliation{2}{Physics Institute, University of Zurich, 8057 Zurich, Switzerland} \\
  \naffiliation{3}{STAR Institute, Quartier Agora - All\'ee du six Ao$\hat{u}$t, 19c B-4000 Li\`ege, Belgium} \\
  \naffiliation{4}{Physik-Department, Technische Universit\"at M\"unchen, James-Franck-Str.~1, 85748 Garching, Germany} \\
}

\date{}
\pubyear{}

\begin{document}

\label{firstpage}
\pagerange{\pageref{firstpage}--\pageref{lastpage}}

\maketitle

\begin{abstract}
  \noindent In the coming years, strong gravitational lens discoveries
  are expected to increase in frequency by two orders of magnitude.
  Lens-modelling techniques are being developed to prepare for the
  coming massive influx of new lens data, and blind tests of lens
  reconstruction with simulated data are needed for validation.  In
  this paper we present a systematic blind study of a sample of 15
  simulated strong gravitational lenses from the EAGLE suite of
  hydrodynamic simulations.  We model these lenses with a free-form
  technique and evaluate reconstructed mass distributions using
  criteria based on shape, orientation, and lensed image
  reconstruction.  Especially useful is a lensing analogue of the
  Roche potential in binary star systems, which we call the {\em
    lensing Roche potential}.  This we introduce in order to factor
  out the well-known problem of steepness or mass-sheet degeneracy.
  Einstein radii are on average well recovered with a relative error
  of ${\sim}5\%$ for quads and ${\sim}25\%$ for doubles; the position angle of
  ellipticity is on average also reproduced well up to $\pm10\degree$,
  but the reconstructed mass maps tend to be too round and too
  shallow.  It is also easy to reproduce the lensed images, but
  optimizing on this criterion does not guarantee better
  reconstruction of the mass distribution.
\end{abstract}


%
\begin{section}{Introduction}\label{sec:Introduction}

  Since the first discovery of a strongly lensing galaxy \citep{Walsh79} these
  systems have been used to study a range of scientific questions: the
  dark-matter (DM) density profile in galaxies and possible substructure
  \citep[see][]{Koopmans03, TreuKoopmans04, Read07, Koopmans09, Auger10,
  Barnabe11, Nierenberg17}, the star formation efficiency and stellar mass
  function \citep[see][]{Leier11, Leier16, Kueng18}, cosmological parameters
  \citep{Paraficz10, Suyu10, Paraficz14, Suyu17}, and the structure of lensed
  active galactic nuclei \citep[see][]{Sluse12, Tomozeiu18, Hutsemekers19}.

  In the coming years many more lens discoveries are anticipated,
  increasing the number of galaxy lenses from $\sim10^3$ to $\sim10^5$
  \citep{Oguri10}.  To prepare for this stream of new gravitational
  lensing data, it is important to have the tools to process
  information efficiently and correctly.  Recent work to this end
  mainly aims to improve the efficiency with which gravitational
  lenses are analysed.  In the verification of lens candidates,
  machine-learning-based projects and citizen science programs have
  been established, working concurrently, as well as complementary, to
  confirm and select interesting objects \citep{DeepLens, Jacobs17,
    SW1, SW2}.  Since the numbers of gravitational lenses will be too
  high for experts to model all by themselves, efforts have been made
  to create crowd-sourced lens-modelling tools such as SpaghettiLens
  by~\cite{Kueng15,SpL-stack}, or automatic modellers like AutoLens
  by~\cite{AutoLens}, and the machine-learning code Ensai
  by~\cite{Hezaveh17}.

  Since all the aforementioned scientific questions require
  reconstruction of the mass distributions of lenses, it is important
  for lens-modelling techniques to be rigorously tested.  It is an
  intrinsic problem of gravitational lensing that many plausible mass
  distributions are able to explain observed data.  This was actually
  pointed out in the very first paper on lens modelling
  \citep{Young81} and is generally known as lensing degeneracy
  \citep[see][]{Saha2000}.  Some degeneracies can be removed by adding
  additional information such as kinematics, but there exist complex
  degeneracies \citep{Saha06, SchneiderSluse14} whose systematic
  effect on modelling has yet to be explored.
  Figure~\ref{fig:degeneracy} illustrates the sometimes surprising
  character of lensing degeneracies.  It is clear that blind tests of
  lens modelling with simulated data are needed.  Such tests have
  already been performed on cluster lenses in the format of a
  challenge in which 10 groups participated and tested their various
  modelling methods \citep{FrontierFieldsTest}.

  \begin{figure}
  \includegraphics[width=0.25\textwidth]{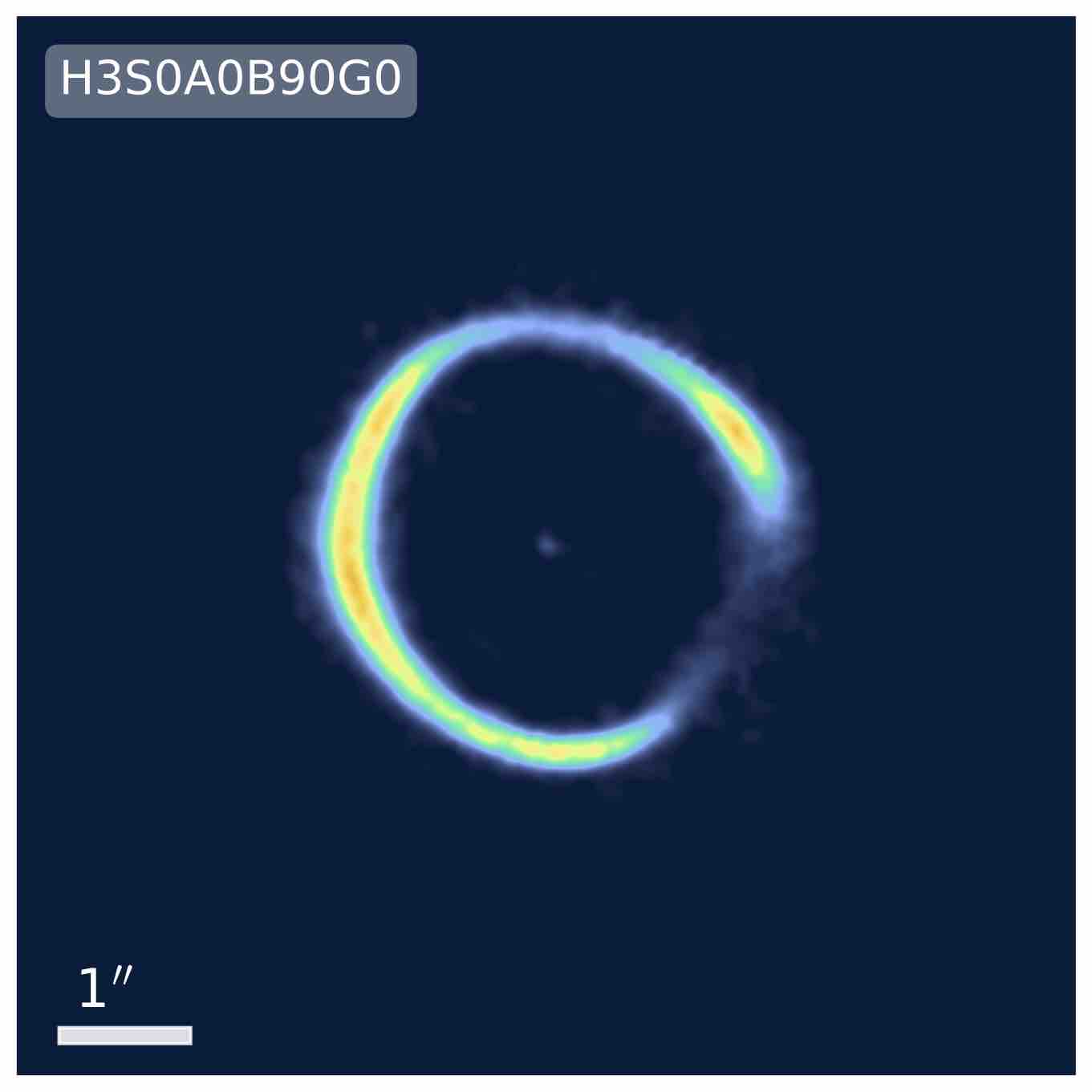}\includegraphics[width=0.25\textwidth]{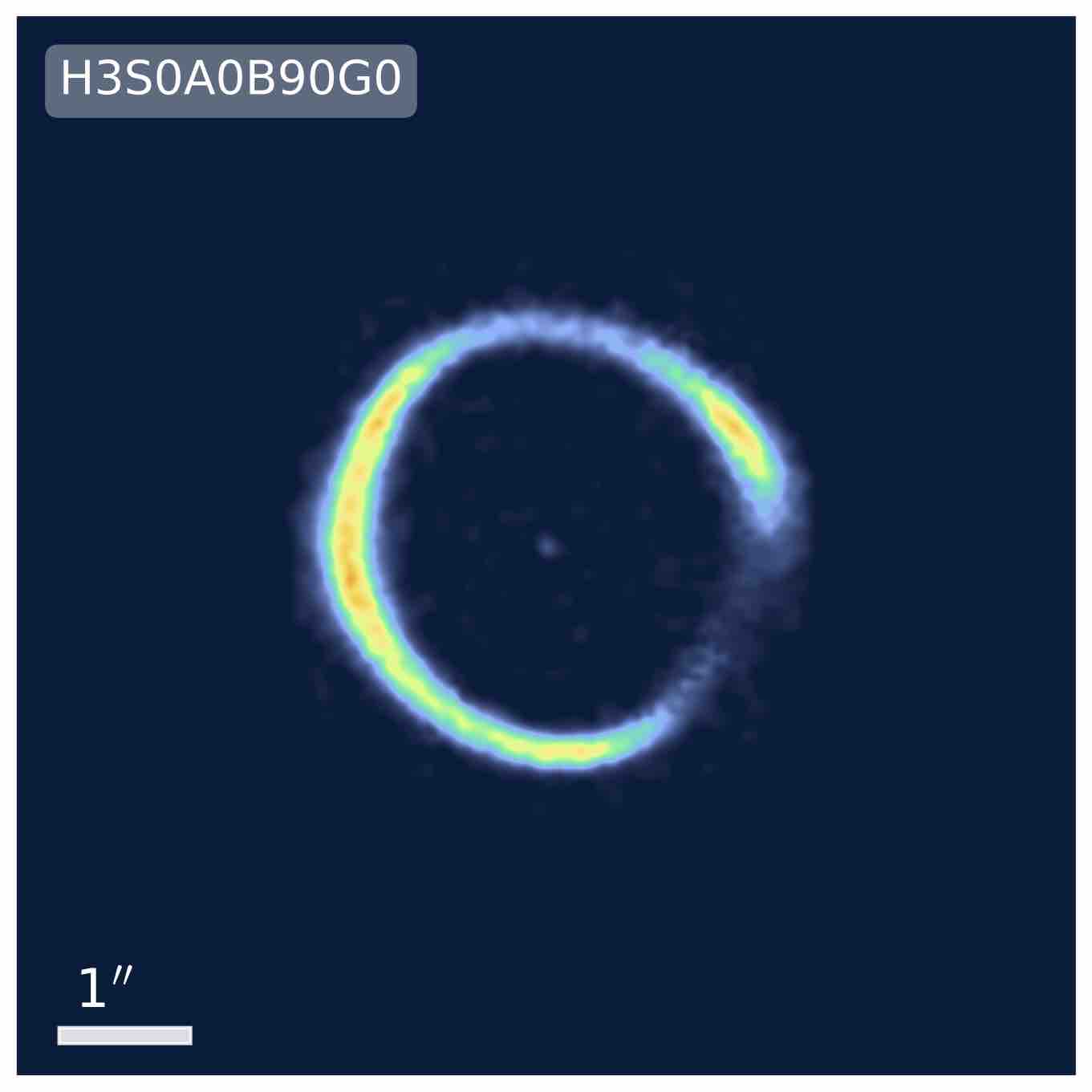}
  \includegraphics[width=0.25\textwidth]{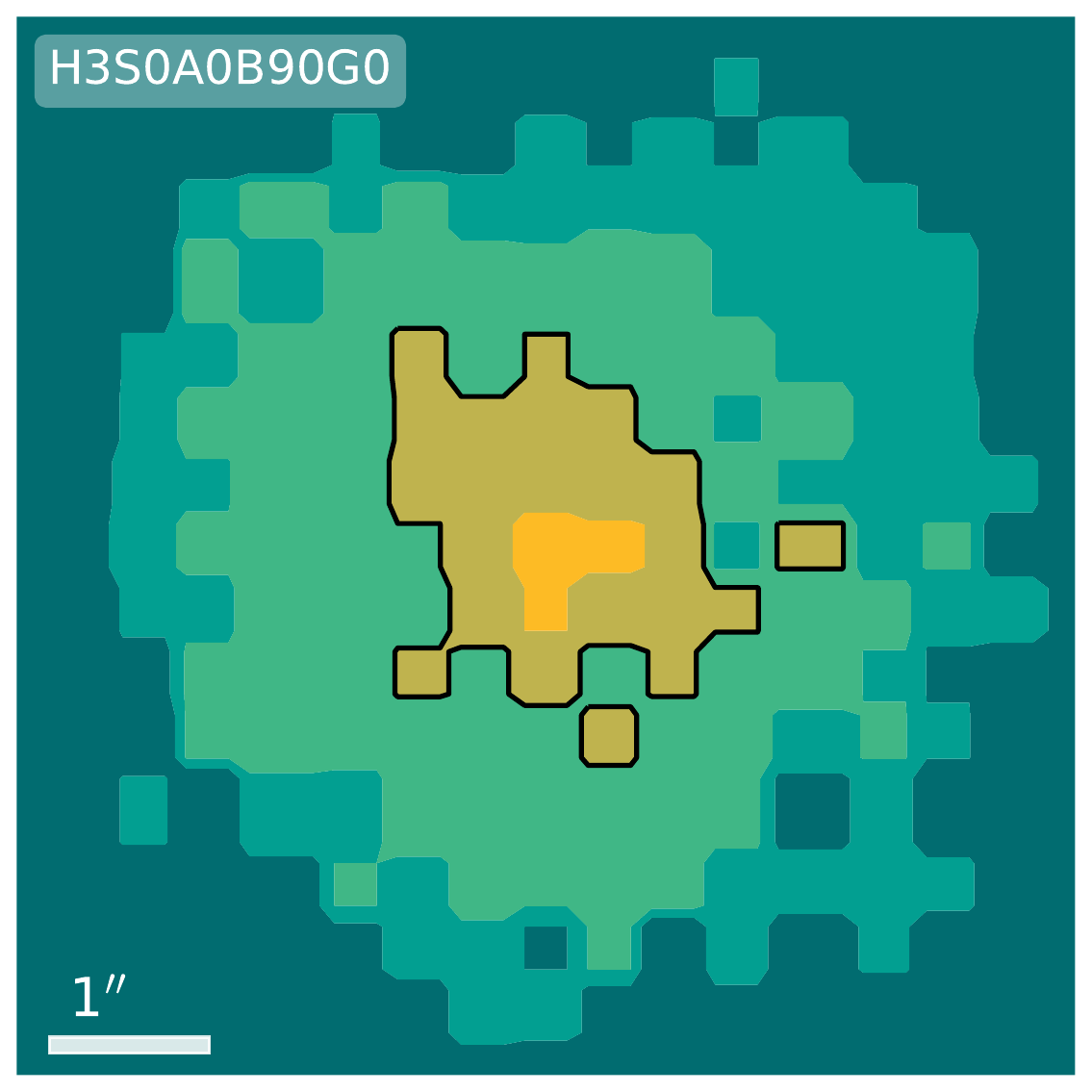}\includegraphics[width=0.25\textwidth]{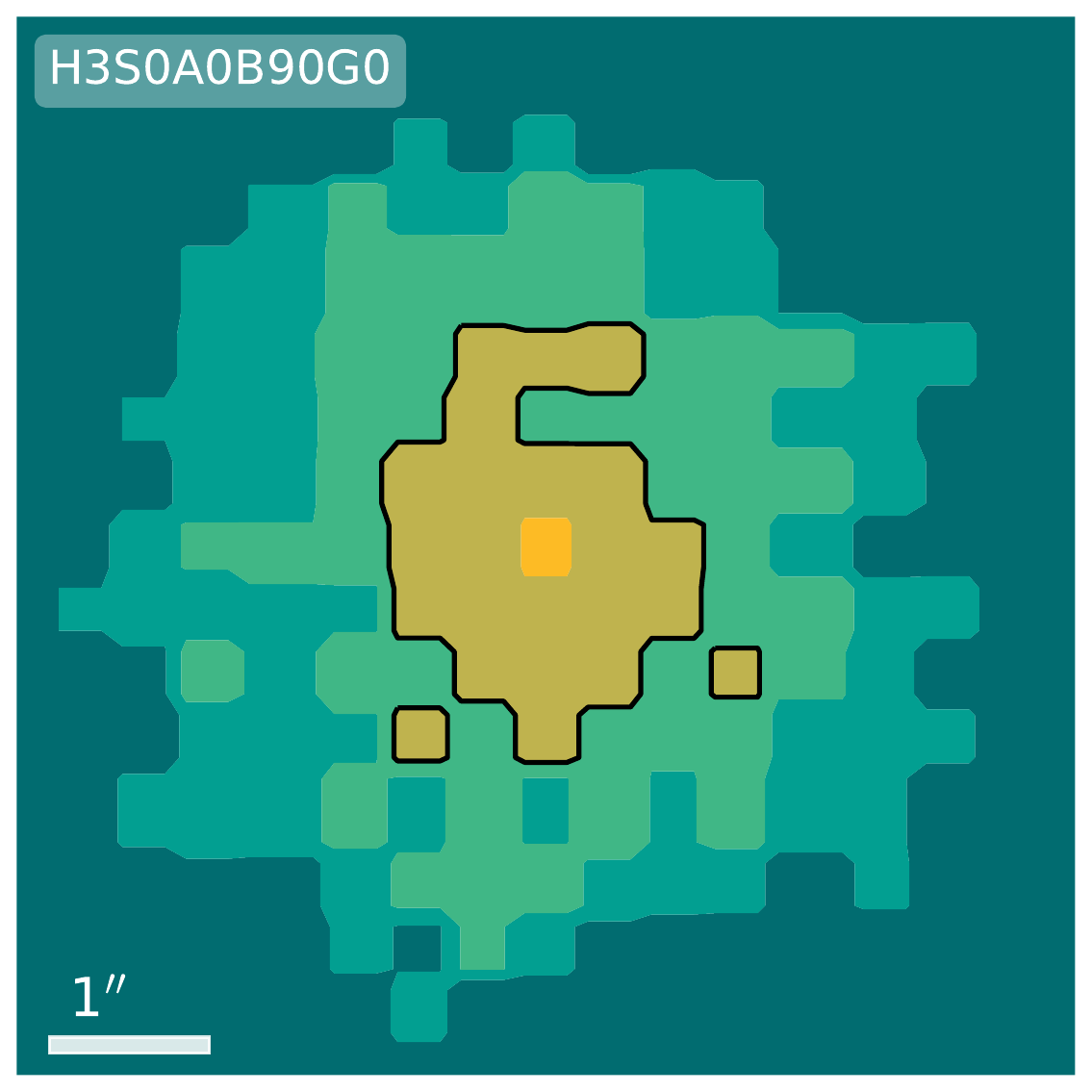}
  \caption{An example of the degeneracy of lens reconstructions.  The
    two lower panels show two different surface mass distributions
    (black lines indicate convergence $\kappa=1$).  The upper panels
    show lensed images produced by these mass maps.  The projected
    images are still very similar to each other, even though the mass
    maps show clear differences in both shape and slope.  This
    illustrates that many different mass distributions may result in
    the same projected image configuration.}\label{fig:degeneracy}
\end{figure}

  The technique of free-form lens reconstruction is a response to the
  problem of lensing degeneracies.  Free-form lens modelling
  deliberately uses underconstrained models and explores lensing
  degeneracies in a high-dimensional parameter space.  The lens mass
  distribution is built out of a superposition of a large number of
  small components, with only few assumptions about the global
  properties of the distribution, such as smoothness, non-negative
  masses, no extra images, or being centrally concentrated.  Early
  free-form models used regularization \citep{Saha97}, but a better
  strategy is to generate a whole ensemble of models which all fit the
  data with a range of different mass distributions
  \citep{pixel2000,PixeLens,Lubini12}.  If the parameter space is
  properly defined and sampled, the possibility is high that the
  \textit{true} solution is contained within the ensemble model.  A
  drawback of such solution sampling strategies is that ensembles
  often also contain physically unrealistic models which are in
  conflict with our current understanding of galaxy evolution.  Such
  unrealistic models might bias the ensemble average, and
  post-processing with further non-linear constraints is then required
  to filter the ensemble.

  This paper explores methods to evaluate and test the quality of ensemble
  models from (free-form) lens-modelling tools, and specifically GLASS
  \citep{GLASS} in a blind study.  In a first phase, we reconstructed 15 lenses
  from the EAGLE suite~\citep[Evolution and Assembly of GaLaxies and their
  Environment;][]{EAGLE} simulated with SEAGLE (Simulating EAGLE LEnses)
  by~\cite{SEAGLE-I}, unaware of the actual mass distribution of the lenses.  We
  investigated the mapping properties between source and lens plane of the
  individual models in the ensemble by creating \textit{synthetic images}, which
  are recovered by using the lensing data to reconstruct the source plane and
  reprojecting back on to the lens plane.  The quality of the data was then
  evaluated by comparing the synthetic images to the original data. Moreover,
  the arrival-time surfaces and mass distribution maps of each model in the
  model ensembles were visually inspected and evidently flawed models were
  filtered out.  In a second phase, three mass distributions were unveiled and
  compared with the models.  The three lenses were chosen to cover most
  distinguishing key properties of all lenses such as doubles, quads, and
  visible maxima.  The reason for only unveiling three lenses in the beginning
  was to be able to improve on the rest of the lens set.  In the third phase,
  all lens reconstructions were revised and finally compared with the simulated
  mass distributions.

  This paper is structured as follows: Section~\ref{sec:Theory} gives
  an overview of the theoretical framework necessary to understand the
  methods detailed in Section~\ref{sec:Methods}.  In
  Section~\ref{sec:Results}. we report on the results obtained during
  the analysis in the different phases, and in
  Section~\ref{sec:Conclusion} the results are discussed and
  summarized.
 
\end{section}

%
\begin{section}{Theory}\label{sec:Theory}

  \subsection{Arrival-time surface}\label{sec:arriv}

  Lensing can be expressed as arrival times of paths from the source to the
  observer which are extremal when the path corresponds to an actual light ray
  according to Fermat's principle.  For a source at $\bm{\beta}$ the arrival
  time for light coming to the observer from position vector $\bm{\theta}$ on
  the sky, is
  \citep[following][]{Blandford86}
  \begin{equation}
    \label{eq:arriv-units}
    t(\bm{\theta}) = \frac{1+z_{L}}{2} \frac{D_{L}D_{S}}{cD_{LS}}(\bm{\theta}-\bm{\beta})^{2} - (1+z_{L})\frac{8\pi G}{c^{3}}\nabla^{-2}\Sigma(\bm{\theta})
  \end{equation}
  where $D_{L}$, $D_{S}$, and $D_{LS}$ are angular diameter distances from
  observer to lens, observer to source, and lens to source, $z_{L}$ is the
  redshift of the lens, $\Sigma(\bm\theta)$ the sky-projected density of the
  lens, and $\nabla^{-2}$ is an inverse Laplacian in 2D.  The
  expression~\eqref{eq:arriv-units} depends, through the distances, on both
  source and lens redshifts.  In cluster lensing, there are typically multiple
  sources at different redshift and sometimes multiple lens redshifts as well.
  In galaxy lensing, it is very rare to have more than a single lens and a
  single source.  Hence, it is convenient to take out the redshift dependence
  and consider a scaled arrival time without the dependence on the redshifts.
  Writing
  \begin{equation}
  \label{eq:scaled_eqs}
  \begin{aligned}
  \Sigma(\bm{\theta}) &= 
  \frac{c^{2}}{4\pi G}\frac{D_{L}D_{S}}{D_{LS}} \, \kappa(\bm{\theta}) \\
  t(\bm{\theta}) &= (1+z_{L}) \frac{D_{L}D_{S}}{cD_{LS}} \, \tau(\bm{\theta})
  \end{aligned}
  \end{equation}
  and discarding the constant $\frac12\beta^{2}$ we have
  \begin{equation}
  \label{eq:arriv}
    \tau(\bm{\theta}) = {\textstyle\frac12}\theta^2
    - 2\nabla^{-2}\kappa(\bm{\theta}) - \bm{\theta}\cdot\bm{\beta}.
  \end{equation}
  The dimensionless surface density $\kappa(\bm\theta)$ is known as the
  convergence.  Through equation~\eqref{eq:scaled_eqs} $\kappa = 1$ corresponds
  to a critical surface density (for circular lenses, this lies within the
  Einstein radius) which characterizes the lens system.  The function
  $\tau(\bm\theta)$ is known as the arrival-time surface. It is not itself
  observable, but still very useful since other relations can be derived from
  it~\citep[see e.g.][]{Courbin02}.  In particular, the lens equation
  \begin{equation} \label{eq:lenseq}
    \nabla\tau(\bm{\theta}) = 0
  \end{equation}
  is just Fermat's principle applied to the arrival time while the inverse
  magnification tensor equals the matrix of second derivatives of
  $\tau(\bm{\theta})$.

  \subsection{Lensing Roche potential}\label{subsec:Roche}

  We now introduce a further insightful quantity.  Let us rewrite the arrival
  time~\eqref{eq:arriv} slightly as
  \begin{equation}
  \begin{aligned} 
    \label{eq:Roche}
    \tau(\bm{\theta}) &= \mathcal{P}(\bm{\theta})
                       - \bm{\theta}\cdot\bm{\beta} \\
    \mathcal{P}(\bm{\theta}) &= {\textstyle\frac12}\theta^2
                              - 2\nabla^{-2}\kappa(\bm{\theta}) \\
                             &= 2\nabla^{-2}(1-\kappa(\bm{\theta})).
  \end{aligned} 
  \end{equation}
  The term $\mathcal{P}$ amounts to solving a 2D Poisson for a potential and
  then adding a centrifugal term.  That is reminiscent of the Roche potential in
  celestial mechanics, and so we will call $\mathcal{P}$ the \textit{lensing
  Roche potential}.  It is like the arrival-time surface without the tilting
  term $\bm{\theta}\cdot\bm{\beta}$, and it represents the arrival-time surface
  for a source at the coordinates origin.  The lensing Roche potential also
  amounts to solving Poisson's equation with $1-\kappa$ as the notional source.

  Consider what happens if we multiply $1-\kappa$ by a constant factor
  $\lambda$.  This changes the steepness of the mass distribution.  Then,
  $\mathcal{P}$ will also get multiplied by $\lambda$.  If $\bm{\beta}$ is
  simultaneously multiplied by $\lambda$, the net result is to multiply $\tau$
  by $\lambda$.  This operation leaves the extremal points unchanged.  That
  means the images are also unchanged, except that the overall magnification
  gets multiplied by $\bm{1}\lambda$ because we have rescaled the whole source
  plane.  This is the steepness degeneracy (also called the mass-sheet
  degeneracy).  In fact, changing $\tau(\bm{\theta})$ in a way that does not
  affect the extremal points will not change the images, so there are really
  infinitely many degeneracies~\citep{Saha2000}.  But the steepness degeneracy
  is the simplest and most severe.  The advantage of $\mathcal{P}$ is that the
  steepness degeneracy only multiplies it by a constant, and does not change its
  shape.  Hence, a normalized lensing Roche potential offers a convenient way of
  comparing different models with the effect of the steepness degeneracy taken
  out.

  \subsection{Lensing mass quadrupole}\label{sec:quadrupole}

  Apart from the Einstein radius, which measures the total mass in the lensing
  region, the shape and orientation of the lensing mass are also of interest.
  We can probe these by considering the 2D inertia tensor
  \begin{equation}
    \label{eq:inertia}
    I = \frac{1}{\pi}
    \left({
        \begin{array}{cc}
          \int \kappa(\bm{\theta})\theta_{x}^{2}\mathrm{d}\bm{\theta} & \int \kappa(\bm{\theta})\theta_{x}\theta_{y}\mathrm{d}\bm{\theta} \\
          \int \kappa(\bm{\theta})\theta_{y}\theta_{x}\mathrm{d}\bm{\theta} & \int \kappa(\bm{\theta})\theta_{y}^{2}\mathrm{d}\bm{\theta} \\
        \end{array}}
    \right)
  \end{equation}
  and its eigenvectors and eigenvalues.  The position angle $\phi$ of the lens
  is given by the angle between one of the eigenvectors of~\eqref{eq:inertia}
  and a unit basis vector.  The semimajor and semiminor axes are given by $a =
  2\sqrt{\lambda_{1}}$ and $b = 2\sqrt{\lambda_{2}}$ where the $\lambda_i$ are
  the eigenvalues.  The ratio $q=b/a$ is a measure of the ellipticity of the
  mass distribution.

  \subsection{Image formation}\label{sec:mappings}

  The lens equation~\eqref{eq:lenseq} defines a mapping from $\bm\beta$ to
  $\bm\theta$.  Let $L(\bm{\theta},\bm{\beta})$ denote that mapping. Then, a
  source-brightness distribution $s(\bm{\beta})$ will produce an image
  brightness distribution
  \begin{equation}
    \label{eq:ideal-mapping}
    d(\bm{\theta}) = \int L(\bm{\theta}, \bm{\beta}) \,
                          s(\bm{\beta}) \,\mathrm{d}\bm{\beta} \,.
  \end{equation}
  This applies if there is perfect resolution on the sky.  In general, there
  will be smearing by the PSF, say $P(\bm{\theta},\bm{\theta}')$.  Hence
  \begin{equation}
    \label{eq:mapping}
    \begin{aligned}
    d(\bm{\theta}) &= \int M(\bm{\theta},\bm{\beta}) \, s(\bm{\beta})
                          \,\mathrm{d}\bm{\beta} \\
    M(\bm{\theta},\bm\beta) &= \int P(\bm{\theta},\bm{\theta}') \,
                               L(\bm{\theta'}, \bm{\beta})
                               \,\mathrm{d}\bm{\theta}' \,.
  \end{aligned}
  \end{equation}
  This mapping is linear in the source brightness $s(\bm\theta)$, but completely
  non-linear in the mass distribution.  Additionally to the smearing effect by
  the point-spread function (PSF), the observed image will contain a foreground
  and noise.  Despite the simplicity of equation \eqref{eq:mapping}, mass
  reconstructions of lenses are far more complex. A source brightness can be
  generated by many different mass distributions. Thus, it is insufficient to
  optimize mass distributions based on source brightness, and in most cases
  special information is needed to break those degeneracies.

\end{section}

\begin{section}{Methods}\label{sec:Methods}

  In this section, we summarize the methodology used in this work.  We first
  describe the SEAGLE pipeline \citep{SEAGLE-I} and how mock lenses were created
  (Section \ref{subsec:SEAGLE}).  We then describe the modelling strategy
  followed in this work with GLASS code (Section \ref{subsec:GLASS}) and
  summarize the source reconstruction process (in Section
  \ref{subsec:synth_method}) and comparison with the lens modelling (in Section
  \ref{subsec:comparison_methods}).

  \subsection{SEAGLE}\label{subsec:SEAGLE}

  SEAGLE is a lens-simulation pipeline based on the EAGLE suite of
  hydrodynamical simulations of the formation of galaxies and supermassive black
  holes in a $\Lambda$-cold dark matter universe \citep{EAGLE, Crain15,
  McAlpine16}.  EAGLE galaxies are in good agreement with observations of the
  star formation rate, passive fraction, Tully--Fischer relation, total stellar
  luminosity of galaxy cluster and colours \citep{Trayford15}, the evolution of
  the galaxy stellar mass function and sizes \citep{Furlong15a, Furlong15b}, and
  rotation curves \citep{Schaller15}. The subgrid physics employed in EAGLE is
  based on that developed for OWLS \citep[OverWhelmingly Large
  Simulations][]{Schaye10}.  In this paper, we have chosen to use the reference
  model L100N1504 which has a maximum proper softening length of 0.7\,kpc
  \citep[see Table 1 in][]{EAGLE, Crain15}.

  We apply the SEAGLE pipeline -- which incorporates the {\tt GLAMER}
  (\citealt{Metcalf14, Petkova14}) ray-tracing code and the parametric
  lens-modelling code {\tt LENSED} (\citealt{Tessore15a,
    Bellagamba17}) -- to selected galaxies from the simulations based
  on their stellar mass and create their DM, stellar, and gas
  surface mass density maps with their corresponding lensed images and
  convergence maps.

  The SEAGLE pipeline automatically identifies and extracts samples of
  (lens) galaxies from the {\tt Friends-Of-Friend} ({\tt FoF})
  catalogues of the EAGLE simulations.  After selecting the galaxy
  identifiers using an initial selection function, we identify the
  desired {\tt GroupNumber}, and {\tt SubGroupNumber} (numbers
  assigned to {\tt FoF} group and subgroup, respectively) and select
  their particles (gas, DM, and stars).  The particles of each galaxy
  from the simulations are rotated in three different axes and
  converted into projected surface density maps (or generally referred
  to as 'mass maps').  The surface density maps are created in units
  of solar masses per pixel on grids of 512$\times$512 pixels (for more
  details of the lensing galaxies, size, and pixel scale of mass maps
  \citep[see table 2 in][]{SEAGLE-I}. Hereby, the entire galaxy and its
  local environment are taken into account \citep[see section 3.2
  in][]{SEAGLE-II}. They are used as input to the ray-tracing lensing
  code {\tt GLAMER} (\citealt{Metcalf14, Petkova14}).  We then choose
  a source redshift for {\tt GLAMER} to convert these mass maps into
  convergence maps.  For each convergence map, the critical curves and
  caustics are calculated to determine where a source has to be placed
  in order to create multiple-lensed images.

  An elliptical S\'{e}rsic brightness profile of the source was used with index
  $n=1$, $\text{apparent magnitude}=23$ in the HST-ACS (Advanced Camera for
  Surveys) F814W filter (AB system) placed at a redshift of $z_{\rm s}=0.6$ to
  mimic SLACS (Sloan Lens ACS Survey) lenses.  The source has an effective
  radius of 0.2\,arcsec, a position angle $ \phi_s=0\degree$ and a constant axis
  ratio $q_s=0.6$.  The pixel scale (0.05\,arcsec), the PSF, and noise
  correspond to an HST-ACS F814W exposure of typically 2400\,s.  The final
  resulting images are exported in standard fits-file format and have sizes of
  161$\times$161 pixels of 8.0 arcsec side length (all parameter values are
  motivated from SLACS lenses, e.~g. \cite{Koopmans06, Newton11}).  We randomly
  choose 15 lensed EAGLE galaxies and their convergence maps (see
  Figures~\ref{fig:data} and~\ref{fig:kappa_true}).

  The nomenclature of the lenses depends on their halo, subhalo, and projection
  catalogue.  A number following 'H' refers to the halo number, 'S' gives the
  subhalo, and letters 'A/B/G' refers to the projection the galaxy has undergone
  in Cartesian coordinates i.e. $\alpha$, $\beta$ and $\gamma$ respectively.
  Although the names of the lenses have a meaning, it is not important in the
  context of this paper, and for simplicity we will rather refer to the lenses
  by their position in the figures.

  \subsection{GLASS}\label{subsec:GLASS}
  The lens modelling was performed with the free-form modelling code
  GLASS \citep{GLASS}.  The strategy behind GLASS \citep[and with its
  precursor {\tt PixeLens}][]{PixeLens} is the following.

  First, $\kappa(\bm\theta)$ is not a simple parametrized form but a free-form
  map made up of a few hundred mass tiles or pixels.  The arrival time
  is then
  \begin{equation}
  \label{eq:free-form-psi}
  \begin{aligned}
    \tau(\bm{\theta}) &= {\textstyle\frac12}\theta^2
                       - \bm{\theta}\cdot\bm{\beta}
    - 2\sum_{n}\kappa_{n}\psi_n(\bm\theta) \\
    \psi_n(\bm\theta) &= \nabla^{-2}Q_n(\bm{\theta})
  \end{aligned}
  \end{equation}
  where $\kappa_n$ is the density of the $n$-th mass tile and
  $Q_n(\bm\theta)$ represents its shape.  Each tile is a square and its
  contribution $\nabla^{-2}Q_n(\bm{\theta})$ can be worked out exactly
  \citep{Abdelsalam}.  The mass tiles are mostly of equal size, but
  smaller tiles are used in the very central region.  We refer to
  $\sim200$ tiles as low resolution and $\sim450$ tiles as high resolution.

  Next, point-like features on the images that correspond to a common
  source are identified using a peak finding algorithm on the lensed images.
  This provides a set of linear equation for
  $\kappa_n$ and $\bm{\beta}$.  These equations have infinitely many solutions,
  from which we sample an ensemble of lens models, according to the
  following priors:
  \begin{enumerate}[leftmargin=*]
  \item $\kappa_n\geq0$,
  \item at each image position, the eigenvalues of $\nabla\nabla\tau(\bm\theta)$
    correspond to minima, saddle points, or maxima in Fermat's principle, as
    specified for that image,
  \item each $\kappa_n$ is limited to twice the average of its neighbours, to
    ensure a reasonably smooth distributions,
  \item the local density gradient is required to point within a specified angle
    (by default $45\degree$ but usually $80\degree$ in this work) of the centre,
  \item azimuthally averaged density profiles must decrease with increasing
    radius,
  \item a constant two-component, external shear is allowed, but each component
    limited to $0.3$.
  \end{enumerate}
  Nominal lens and source redshifts of $z_{L}=0.5$ and $z_{S}=3.0$ and a
  concordance cosmology with
  $(\Omega_{m}, \Omega_{\Lambda}, H_{0}^{-1})= (0.3, 0.7, 14\,{\rm Gyr})$ were
  assumed, as this information was unknown during the blind phase.
  These values do not enter the results however, as the subsequent
  analysis is entirely in terms of $\kappa$.

  \subsection{Source reconstruction and synthetic images}\label{subsec:synth_method}

  GLASS treats a source as a point on the source plane that can map on to
  multiple points on the image plane.  Its solutions consist of ensembles of
  convergence maps which are constrained to reproduce the given image positions
  from some inferred source position.  In reality, we have extended images from
  extended sources.  Extended sources can be emulated by simply using multiple
  point-like features.  For detailed image reconstruction, however, we use a new
  strategy, an extension of the method used by~\cite{Kueng18}, involving
  post-processing the model ensemble from GLASS as follows.

  For each model in the GLASS ensemble, we compute a discrete version
  $M_{\bm{\theta\beta}}$ of the lensing and PSF-smearing map
  $M(\bm\theta,\bm\beta)$ in equation~\eqref{eq:mapping}, by considering
  $\bm\theta$ values at pixel locations.  We then solve for the
  source-brightness distribution $S_{\bm\beta}$ such that the synthetic image
  \begin{equation} \label{eq:matrix-mapping}
  D_{\bm\theta} = \sum_{\bm\beta} M_{\bm{\theta\beta}} S_{\bm\beta}
  \end{equation}
  best fits the observed brightness $D^{\text{obs}}_{\bm\theta}$ in a
  least-squares sense.  That is to say, we minimize
  \begin{equation}
    \label{eq:chi2}
    \chi^{2} = \sum_{\bm\theta}
    \frac{\left(\sum_{\bm\beta}M_{\bm{\theta\beta}}\,S_{\bm\beta}
                - D^{\text{obs}}_{\bm\theta}\right)^{2}}
    {\sigma^2_{\bm\theta}}
  \end{equation}
  for each model in the GLASS ensemble.  Once the source-brightness
  distribution has been fitted, it can be reinserted
  into~\eqref{eq:matrix-mapping} to generate a synthetic image.

  This procedure is formally similar to lens inversion methods previously used
  by \cite{Warren03, Dye05, Brewer06, Suyu06, Vegetti09, Tagore14}, and many
  others.  In their case, the extended lens images help constrain the
  source-brightness model and the parametric lens model simultaneously.  For
  such methods, equation~\eqref{eq:chi2} is usually supplemented with
  regularization terms to penalize noisy solutions.  A problem which
  \cite{Warren03} already pointed out is that regularization produces too smooth
  source-brightness distributions and lead to a bias in the model.  Moreover, it
  was found that mass models are rather insensitive to regularizations if an
  optimal source pixel size is chosen.  Here, the methodology is different. We
  use equation~\eqref{eq:chi2} to solve for synthetic images with already
  modelled mass maps and thereby test our models' mapping properties.  Since we
  are only interested in the lensed images, regularizations are not necessary.

  The uncertainty of an observation corresponding to $\sigma_{\bm\theta}$ in
  equation~\eqref{eq:chi2} generally has several sources, and depends on the
  optical devices in use.  The simplest to model is the photon noise
  \begin{equation}
    \label{eq:noise}
    \sigma_{\bm\theta}^{2} = g^{-1} D^{\text{obs}}_{\bm\theta}
  \end{equation}
  where $g$ is the gain or counts per photon.  We also assumed a
  further uniform noise field to mimic other noise sources.

  No luminosity and kinematic information about the lensing galaxy was included
  in the data, nor was any information about the light distribution of the
  unlensed source.  Not having light from the lensing galaxy has one benefit,
  namely not polluting the lensed image, but also has disadvantage of removing
  potentially useful information about the lens.

  The above is implemented in \verb|python| so to easily interface
  with GLASS, but uses optimized parts written in \verb|cython|
  \citep[see][]{cython} and \verb|C|.  To solve the linear inverse
  problem it relies on methods provided in the module
  \verb|scipy.sparse| and \verb|scipy.sparse.linalg|
  \citep[see][]{scipy}.  Computationally, the most expensive operation
  is the construction of the sparse matrix $L_{\bm{\theta\beta}}$ which when
  multiplied with the PSF yields $M_{\bm{\theta\beta}}$, because matrix
  multiplication routines are highly optimized in most frameworks.

  \subsection{Model comparison}\label{subsec:comparison_methods}

  Individual models of the lens reconstruction ensemble were compared
  to the SEAGLE lens models to evaluate their goodness of fit.

  First, the resemblance of the convergence maps was investigated.  As
  gravitational lensing is foremost determined by the total mass
  content of the lens, the ensemble models' Einstein radii --- which are
  a measure of the total mass contained within --- were compared to the
  SEAGLE models.  By definition, the scale of the Einstein radius is
  set where the radial profile of the convergence crosses $\kappa = 1$,
  thus for non-circular lenses we adapt the definition of a notional
  Einstein radius where $\langle \kappa\rangle_{\mathrm{R}_{\mathrm{E}}} = 1$.
 
  Additionally, the shape and orientation of the lens were analysed.  The
  inertia ellipse's semiminor to semimajor axial ratio $q$ acts as a shape
  parameter and was determined with equation~\eqref{eq:inertia}.  The
  orientation was determined by the position angle $\phi$ as described in
  Section~\ref{sec:quadrupole}.  It can be determined up to an ambiguity of
  $\pm\pi$, provided the lens is not perfectly round (which happens in the limit
  of $q \rightarrow 1$).  To combat this degeneracy, we define a complex
  ellipticity
  \begin{equation}
    \label{eq:epsilon}
    \epsilon = \frac{1-q}{1+q}\,e^{2i\phi}
  \end{equation}
  which combines the ellipticity and the position angle
  following~\cite{SEAGLE-I}.  With this definition, round lenses will fall
  closer to the origin no matter what position angle, whereas more elliptical
  lenses move away from the origin at an angle given by the position angle.
  
  As mentioned before, the relation between convergence and synthetic image is
  degenerate.  This means, if the convergence exactly matches the 'true'
  convergence, the synthetic image and the observed data will be
  indistinguishable.  Slight changes in the convergence however do not linearly
  translate to the synthetic image, and vice versa.  A straight-forward
  comparison of the model's convergence to the 'true' convergence map is thus
  insufficient.  We therefore use the Roche potential as in
  equation~\eqref{eq:Roche} as a basis of comparison.  The best-fitting model in
  the ensemble can be found with the maximum modulus of the inner product of the
  modelled lensing Roche potential $\mathcal{P}^{\text{mod}}$ and the lensing
  Roche potential of the original convergence map $\mathcal{P}^{\text{orig}}$.
  \begin{equation}
    \label{eq:Roche_scalar}
    \max \langle P^{\text{orig}}, P^{\text{mod}}\rangle = \max \left| \frac{\int \mathcal{P}^{\text{orig}}(\bm{\theta}) \mathcal{P}^{\text{mod}}(\bm{\theta}) \mathrm{d}\bm{\theta}}{\sqrt{\int \left|\mathcal{P}^{\text{orig}}\right|^{2}\mathrm{d}\bm{\theta}} \sqrt{\int \left|\mathcal{P}^{\text{mod}}\right|^{2}\mathrm{d}\bm{\theta}}} \right|
  \end{equation}

  Since the Roche potentials are independent of the source position and the
  inner product is normalized, the mass-sheet degeneracy is completely
  eliminated.  The mass-sheet degeneracy not only affects a mass distribution's
  overall amplitude by adding (or subtracting) an arbitrary number of mass
  sheets of critical density, but more importantly also changes its scale and
  steepness.  This means, even though a model's convergence profile might not
  show perfect match in its slope, it can very well be a valid solution.

\end{section}

\begin{section}{Results}\label{sec:Results}

  The sample of 15 SEAGLE lenses is described in Section~\ref{subsec:SEAGLE} and
  shown in Figure~\ref{fig:data}.  These lensed images were prepared by SM.  PD,
  in consultation with JC and PS, reconstructed the lenses from these data.
  Besides the mass centroid of the lensing galaxy and the PSF used to blur the
  lensed images, no other information about the lenses was provided to the
  modellers.

  The lens reconstructions were done in two phases.  In the first phase, three
  lenses were unblinded early, in order to test and improve the pipeline.  In
  the second phase, PD modelled the other 12 lenses unaware of the truth held by
  SM.

  The lenses in the top row were chosen for the first phase because of their
  diverse lensed image configuration.  As shown in Figure~\ref{fig:data}, the
  lens on the left is a quad lens with an almost ring-like image configuration,
  which is present in most regular quads and doubles in the sample, the lens in
  the middle is a double with a wide arc image and a less extended image
  slightly closer to the lens, and the lens on the right is one of five systems
  presenting a fifth non-demagnified image, which is always a maximum of the
  arrival time.  This subset covers the most typical lens properties in the
  entire sample (from a lens modeller's perspective).  After the comparison of
  the three lens models with their actual convergence, the lens-modelling
  procedure was adjusted to be better prepared for the second phase.

  In the second phase, the remaining lenses were reconstructed with the actual
  convergence maps veiled.  The latter were then revealed for comparison, but no
  further changes to the models were allowed.

  It is important to note that our goal during the lens reconstruction was not
  simply to optimally reproduce the lenses images, but to find an ensemble
  sampling the possible mass maps.

  \subsection{Lens reconstructions}\label{subsec:arriv}

  The lens modelling directly yields ensembles of $\kappa$ maps for the lenses.
  Synthetic images for each mass map in the ensemble were constructed and a
  $\chi^{2}$ was evaluated for each image according to equation~\eqref{eq:chi2}.
  In each ensemble at least half of all synthetic images gave bad fits to the
  image data, however for 12 of the 15 lenses there was at least one synthetic
  image which fit the data well.  Figure~\ref{fig:synth} shows the synthetic
  image with the minimal $\chi^{2}$ in the ensemble, for each lens.  As
  described in Section~\ref{subsec:synth_method}, the synthetic images were
  constructed by projecting the data from the image plane on to the source plane
  using the model to calculate the deflection angles, and reprojecting back on
  to the image plane.  Interestingly, and against our initial expectations,
  low-resolution models often seem to produce better synthetic images.  The
  reduced $\chi^{2}$ values never get to 1, indicating that the best fits are
  dominated by systematic errors in the model fitting. The high-resolution
  images have reduced $\chi^{2}$ of $\sim2.5$, whereas the low-resolution images
  have values as low as 1.7.

  Although it is sometimes possible to spot irregularities and unphysical
  features in the mass maps from simple inspection, it is generally easier to
  review the contours of the derived arrival-time
  surface~\eqref{eq:free-form-psi}.  Arrival-time surfaces become very intuitive
  to interpret once the saddle-point contours have been drawn, as they already
  schematically resemble lensed arcs and can visually be compared to the image
  data.  Figure~\ref{fig:arriv} shows arrival-time surfaces of the
  ensemble-average models of each lens.  Saddle-point contours are indicated in
  black, and image position constraints for minima, saddle-points, and maxima in
  blue, purple, and red respectively.  The models by construction reproduce
  point-like image features at the correct positions.  But they can also show
  additional spurious images, easily identifiable as local extrema.  Ensemble
  averages tend to suppress these spurious images.  The arrival-time surfaces
  from GLASS are highly sensitive to the image positions.  In our experience,
  good models produce nice smooth-looking arrival-time surface, but so do some
  bad models; whereas ragged-looking arrival-time surfaces invariable indicate
  bad models.  Slight shifts in the image positions might result in
  significantly different arrival-time surfaces and mass maps.  The fact that
  the lenses have extended, and sometimes almost ring-shaped images, aggravates
  this difficulty.  As the time delays were unknown, the parity of the extrema
  was uncertain as well.  In some cases, a trial-and-error approach had to be
  adopted until a credible model was obtained, with others an educated guess
  could be made based on the distances of the images to the lens and their
  brightness.

  During the reconstructions, the prior parameters in GLASS were tweaked to
  obtain a satisfactory arrival-time surface of the ensemble average in each
  case.  Additionally, we filtered out a percentage of the worst $\kappa$ maps
  from the distribution of $\chi^{2}$, and by mere construction improved the
  ensemble average's synthetic image.  This gave considerable improvements for
  ensembles with a wide variety in its models, however only slight changes for
  ensembles with a less diverse set.

  \subsection{Convergence map comparison}

  The $\kappa$ maps used by SM to generate the lenses are shown in
  Figure~\ref{fig:kappa_true}.  Apart from the top row, these were hidden during
  the modelling process.  The modelling generated ensembles of $\kappa$ maps,
  and the ensemble averages are shown in Figure~\ref{fig:kappa}.  The modelled
  convergence maps show the mass-tile structure of the free-form method, whereas
  the actual maps have higher pixel resolution.  A direct comparison of the
  convergences was expected to be insufficient due to the well-known problem of
  degeneracies.  The numerically best-fitting convergence map of an ensemble
  according to
  \begin{equation}
    \min\limits_{\text{model}} \sum_{i}{(\kappa^{\text{truth}}_{i} - \kappa^{\text{model}}_{i})}^{2}
  \end{equation}
  produced bad or mediocre synthetic images in all lens models, which further
  confirmed our suspicion.  We visually inspected all ensembles each with 1000
  individual models. In some ensembles, the ellipticities and position angles
  had little spread and were definitive, but in others, the models showed an
  ambiguity of $\pm \pi$ in their position angles.  We therefore compared the
  radial profiles and ellipticities, as follows.

  Figure~\ref{fig:kappa_profile} shows the average enclosed $\kappa$ as a
  function of radial distance from lens centre.  The formal Einstein
  $\mathrm{R}_{\mathrm{E}}$ corresponds to $\langle
  \kappa\rangle_{\mathrm{R}_{\mathrm{E}}} = 1$.  The Einstein radii are well
  recovered and have a median relative error of 4.3 per cent over the entire set
  of quads; for some quads, the Einstein radius had an error as low as 1.0 per
  cent, and was slightly overestimated for others with maximally 15.7 per cent,
  but in all cases the errors were smaller than the pixel sizes. The Einstein
  radii for double systems were less accurate with an average error of 24.8 per
  cent.  But even so, the profiles are systematically too shallow.  It is
  interesting, however, that the model profiles (green-yellow bands) could be
  brought much closer to the correct profiles (red curves) by multiplying
  $1-\kappa$ by a constant.  This transformation is nothing but the steepness
  degeneracy, which does not affect the images.

  Figure~\ref{fig:ellipticity} compares the complex
  ellipticity~\eqref{eq:epsilon} for the model ensembles with the SEAGLE values.
  Without exception, the lens models tend to be too round.  This makes it more
  difficult to determine a position angle for the models.  Nevertheless, in all
  but two cases the position angle was roughly recovered with a median error of
  $\pm9.4\degree$.  This is also observable by comparing corresponding maps in
  Figure~\ref{fig:kappa_true} and Figure~\ref{fig:kappa}.

  The problematic lenses are in the bottom row, especially the two in the middle
  and right-hand panels.  In both cases, the position angles are off by almost
  90$\degree$.  This is evident in Figure~\ref{fig:ellipticity} as the
  ellipticity of the ensemble model is in the quadrant opposite to the one of
  the SEAGLE simulation.  This is also observable by directly comparing the
  orientation of the galaxies in Figure~\ref{fig:kappa_true} and
  Figure~\ref{fig:kappa}.  After comparing the arrival-time surfaces and
  convergences, we suspect the image order is most likely wrong.  In both cases,
  the arc to the north should have been diagnosed as arriving before the
  counterimage to the south, whereas in the models the opposite was done.  The
  estimated semimajor and semiminor axes which appear in the complex ellipticity
  only as a ratio, also were generally too high compared to the SEAGLE
  simulations.  This indicates that the lens ensemble models have more mass at
  higher radii from the centre relative to the total mass, as the semimajor and
  semiminor axes were estimated with the inertia ellipses of the convergences.
  This means, as we have already seen, that the models tend to be too shallow
  across the board.

  \subsection{Roche potential comparison}

  After unblinding the actual convergence maps of the lenses in the top row, it
  became clear that a further comparison was desirable which quantified the
  goodness of the lens reconstruction with the effect of the steepness
  degeneracy taken out.  We then formulated and calculated the lensing Roche
  potential.  This is just the arrival time from
  equation~\eqref{eq:free-form-psi} without the $\bm{\theta}\cdot\bm{\beta}$
  term and an arbitrary additive constant.  Furthermore, the effect of the
  steepness degeneracy corresponds to an arbitrary multiplicative factor, as
  explained in Section~\ref{subsec:Roche}.  Thus, we are free to subtract the
  potential's value at the centre from the entire map and to normalize to the
  negative of its minimal value.  This way the lowest minimum has the value $-1$
  and the map centre the value $0$, and takes out the effect of the steepness
  degeneracy.  The results from the actual convergence maps and from the lens
  models are shown in Figure~\ref{fig:roche_true} and Figure~\ref{fig:roche}
  respectively.  These figures show contoured regions of equal level of the
  Roche potentials as in equation~\eqref{eq:Roche}.

  In a rough visual comparison, most lens models --- except the already
  mentioned lenses in the bottom row --- show the same main features and
  morphologically appear to agree with their SEAGLE counterpart.  In
  particular, the position and shape of minima and the lens'
  orientation match quite well.  For some quad systems such as the
  left one in the second-last row, there seems to be a tilt in the
  model which is suppressing one of the minima, which is not evident
  in the corresponding SEAGLE simulation.  Similarly, some saddle
  regions of the models seem to have switched amplitudes causing
  lemniscates to wrap around the central maximum from a different side
  compared to the SEAGLE simulations.

  A quantitative comparison of the Roche potentials was done by
  evaluating the scalar product \eqref{eq:Roche_scalar}.  The black
  circles in Figure~\ref{fig:roche_true} and~\ref{fig:roche} indicate
  the radii within which integrals in the scalar product were
  evaluated.  It was necessary to choose a radius smaller than the map
  radius as otherwise the scalar product would have been dominated by
  differences on the edges of the image plane.  Using a small
  integration radius instead, we observed that differences in the
  shape of the central region and regions near the saddle-contours of
  the potential dominate the scalar product.

  Filtering out models from the ensemble which give low scalar product
  values will by construction yield an overall improved ensemble
  average, since the fraction of bad models shrinks.

  It is now interesting to put in contrast the synthetic image and the Roche
  potential as criteria.  This means comparing the distribution of $\chi^{2}$
  from the synthetic images and the distribution of scalar products of the Roche
  potentials $\langle P^{\text{orig}}, P^{\text{mod}}\rangle$ within an
  ensemble.  This comparison is illustrated by the hex-binned plots in
  Figure~\ref{fig:chi2_VS_roche_scalar}.  Just as we expect a filtering of the
  ensemble in which only models with low $\chi^{2}$ of the synthetic image are
  kept to improve the ensemble overall, filtering according to the Roche
  potential is expected to elevate the quality of ensembles as well.  Hence we
  expect both methods to anticorrelate, meaning, models with a low $\chi^{2}$
  should to have a scalar product of the Roche potential with a value close to
  1, whereas higher $\chi^{2}$ should have a low scalar product.  At first
  glance, Figure~\ref{fig:chi2_VS_roche_scalar} neither disproves this
  assumption, nor conclusively confirms it.  None of the lenses clearly display
  tendencies towards anticorrelation between $\chi^{2}$ and $\langle
  P^{\text{orig}}, P^{\text{mod}}\rangle$.  However, it seems that mostly models
  with a diverse ensemble show an anticorrelative trend, whereas models with an
  already overall high-quality ensemble distribute rather uniformly in that
  parameter space.  It is also interesting that some lenses, e.g. the lens in
  the middle row to the right (H30S0A0B90G0), show a wide spread in the scalar
  product of the Roche potentials, but a very low spread in $\chi^{2}$. This
  means that even though the shapes of the mass maps differ widely, the mass
  maps produce similar quality of synthetic images.  This implies that filtering
  should improve broad ensemble models, however there is a point when optimizing
  an ensemble model does not change the model anymore, at least globally.  This
  is probably due to the fact that those models seem to agree on the global
  structure of the convergence and continue fitting substructures.

  An example of this is also shown Figure~\ref{fig:filtered}.  The differences
  are minor, but reducing the ensemble to only 100 models with the lowest
  $\chi^{2}$ visibly improves the synthetic image of the ensemble-averaged
  model, which was expected as the $\chi^{2}$ is a direct measure of the
  synthetic images.  However, trying to improve that ensemble again by filtering
  out the worst scalar products of the Roche potentials does not clearly improve
  upon the already filtered ensemble anymore.  This can be explained by the fact
  that in the end synthetic images are a local, non-linear measure of the mass
  whereas the Roche potential is a direct measure of the global mass model.

\end{section}

\begin{section}{Conclusion}\label{sec:Conclusion}

  We need to prepare for the tens of thousands of strong gravitational lenses
  expected to be discovered with the next generation of wide-field telescopes:
  \textit{Euclid} and \textit{WFIRST (Wide-Field Infrared Survey Telescope)} in
  space and \textit{LSST (Large Synoptic Survey Telescope)} on the ground.
  Lately, many efforts have been made to prepare for this data flood, from
  automatic identification of lens candidates with the use of machine learning,
  to automatic modelling of lenses with novel codes.  While this paper does not
  address the problem of scalability for future big data sets, it demonstrates
  the necessity of such blind test, particularly for lens-modelling tools which
  were designed to scale up to thousands of lenses.  It is still unclear how
  much we can trust the resulting lens models of such pipelines and what aspects
  of a lens model are robust against degeneracies.  The only way to objectively
  test lens reconstruction techniques and avoid confirmation bias are blind
  tests with realistically simulated data.

  In this work, we have used a sample of 15 simulated strong lenses from the
  state-of-the-art hydrodynamical simulation of EAGLE and modelled them blind
  with the GLASS code.  We introduced a new lensing potential, the 'Roche
  potential' and showed that using this in free-form modelling we are
  successfully able to reconstruct the lensing systems without the mass-sheet or
  steepness degeneracy.

  General properties like extended image information, Einstein radii,
  and shape of the convergence have been investigated and compared to
  the actual, subsequently unveiled convergence. Thereby, the
  following results stood out:
  \begin{itemize}[leftmargin=*]
  \item It was straightforward to find models in the ensemble with
    good fitting synthetic images.  Low-resolution models actually
    tended to perform better in this respect.  It appears that
    high-resolution models start fitting substructures which
    predominantly impact the synthetic images.  Also, low-resolution
    images produce smoother source images, which might also improve
    fits to the images.  This can be seen when comparing
    Figure~\ref{fig:degeneracy} which shows synthetic images
    constructed with low-resolution models and Figure~\ref{fig:synth}
    which displays the image reconstructions from high-resolution
    models.
  \item The Einstein radii were recovered quite well for most lenses,
    in some cases they were slightly overestimated.
  \item The position angles were also approximately in agreement,
    except for two lenses.  In those two cases, the problems were
    traced back to having chosen the wrong image order in the lens
    reconstruction.
  \item The ellipticities of the mass maps were generally too round.
  \item The radially averaged convergence profiles were all found to
    be too shallow.  However, this is an effect of the steepness
    degeneracy and could be resolved by multiplying the $1-\kappa$ surface
    by an arbitrary factor.
  \end{itemize}
  Even when all those properties are approximately recovered by a lens model, it
  was demonstrated that this does not necessarily mean the actual distribution
  has been found.

  The reason why reconstructing lenses is so difficult lies in the many kinds of
  degeneracies that affect lens models.  The most important of these is the
  steepness degeneracy (also called the mass-sheet degeneracy).  The novel
  concept of a lensing Roche potential is introduced to remove the effect of the
  steepness degeneracy, that is, to compare models which have been differently
  affected by the steepness degeneracy.  The scalar product of two Roche
  potentials gives a true measure of how alike models are to each other whilst
  ignoring the steepness degeneracy.  A good match between model and truth
  simply means that up to an arbitrary rescaling of $1-\kappa$ the shape of
  $\kappa$ agrees well with the truth.  In the cases where the scalar product of
  the Roche potentials is far from 1, we have a bad match, meaning the actual
  shape of $\kappa$ has not been recovered well enough.

  We demonstrated that filtering out models where the synthetic images
  have the highest $\chi^{2}$ seems to improve ensembles up to a point.
  However, once the global properties of the lens have been modelled,
  further optimization of the synthetic images does not improve fits
  to substructures and other local properties.  This means, even if
  synthetic images from a model ensemble might show only minor
  variation, the underlying convergence might have major differences
  in comparison.  This shows that the scalar product of Roche
  potentials is a better criterion to determine a model's quality when
  mass distributions are of primary interest.

  In summary, this study not only confirms the well-known characteristic of lens
  modelling, that the Einstein radius of the mass distribution is generally well
  reproduced whereas its steepness is not, but also highlights additional
  points:
  \begin{itemize}[leftmargin=*]
  \item To mitigate the problem of unbroken degeneracies, we recommend producing
    an ensemble of many solutions rather than a single model.
  \item While synthetic images represent a convenient visual diagnostic, they
    are useful only up to a point.  Once a synthetic image is reasonably good
    (and well before the formal $\chi^2$ criterion is achieved) the goodness of
    the synthetic image does not correlate with the goodness of the lens model.
    In other words, even if the model reproduces lensed images which resemble
    the data rather nicely, the mass map might still be far from the truth.
  \item Careful attention should be paid to the parity and time order of the
    images, because getting these incorrect, results in poor models.
  \item When testing lens-modelling tools against well-known lenses or
    simulations, the scalar product of the Roche potentials gives a measure of
    the shape similarity of the mass models which is robust against
    degeneracies. If this scalar product indicates a bad match, a recalibration
    of the lens-modelling tool might be needed.
  \end{itemize}

\end{section}


%
\section*{Funding statement}
PD acknowledges support from the Swiss National Science Foundation.
SM acknowledges the funding from the European Research Council (ERC)
under the EU’s Horizon 2020 research and innovation programme
(COSMICLENS;\@ grant agreement no.~787886).

%

%
\bibliographystyle{mnras}
\bibliography{refs}

%
%
\def\pwidth{.33\textwidth}
\def\pheight{.19\textheight}
\newcommand{\SEAGLEmosaic}[2]{
  \centering
  \includegraphics[#2]{imgs/#1/#1_H3S0A0B90G0}\includegraphics[#2]{imgs/#1/#1_H10S0A0B90G0}\includegraphics[#2]{imgs/#1/#1_H36S0A0B90G0}
  \includegraphics[#2]{imgs/#1/#1_H2S2A0B90G0}\includegraphics[#2]{imgs/#1/#1_H3S1A0B90G0}\includegraphics[#2]{imgs/#1/#1_H2S1A0B90G0}
  \includegraphics[#2]{imgs/#1/#1_H160S0A90B0G0}\includegraphics[#2]{imgs/#1/#1_H4S3A0B0G90}\includegraphics[#2]{imgs/#1/#1_H30S0A0B90G0}
  \includegraphics[#2]{imgs/#1/#1_H13S0A0B90G0}\includegraphics[#2]{imgs/#1/#1_H2S7A0B90G0}\includegraphics[#2]{imgs/#1/#1_H1S0A0B90G0}
  \includegraphics[#2]{imgs/#1/#1_H1S1A0B90G0}\includegraphics[#2]{imgs/#1/#1_H23S0A0B90G0}\includegraphics[#2]{imgs/#1/#1_H234S0A0B90G0}
}

\begin{section}{Figures}\label{sec:Figures}

  This section contains all figures which are referenced in previous sections.

  \begin{figure*}
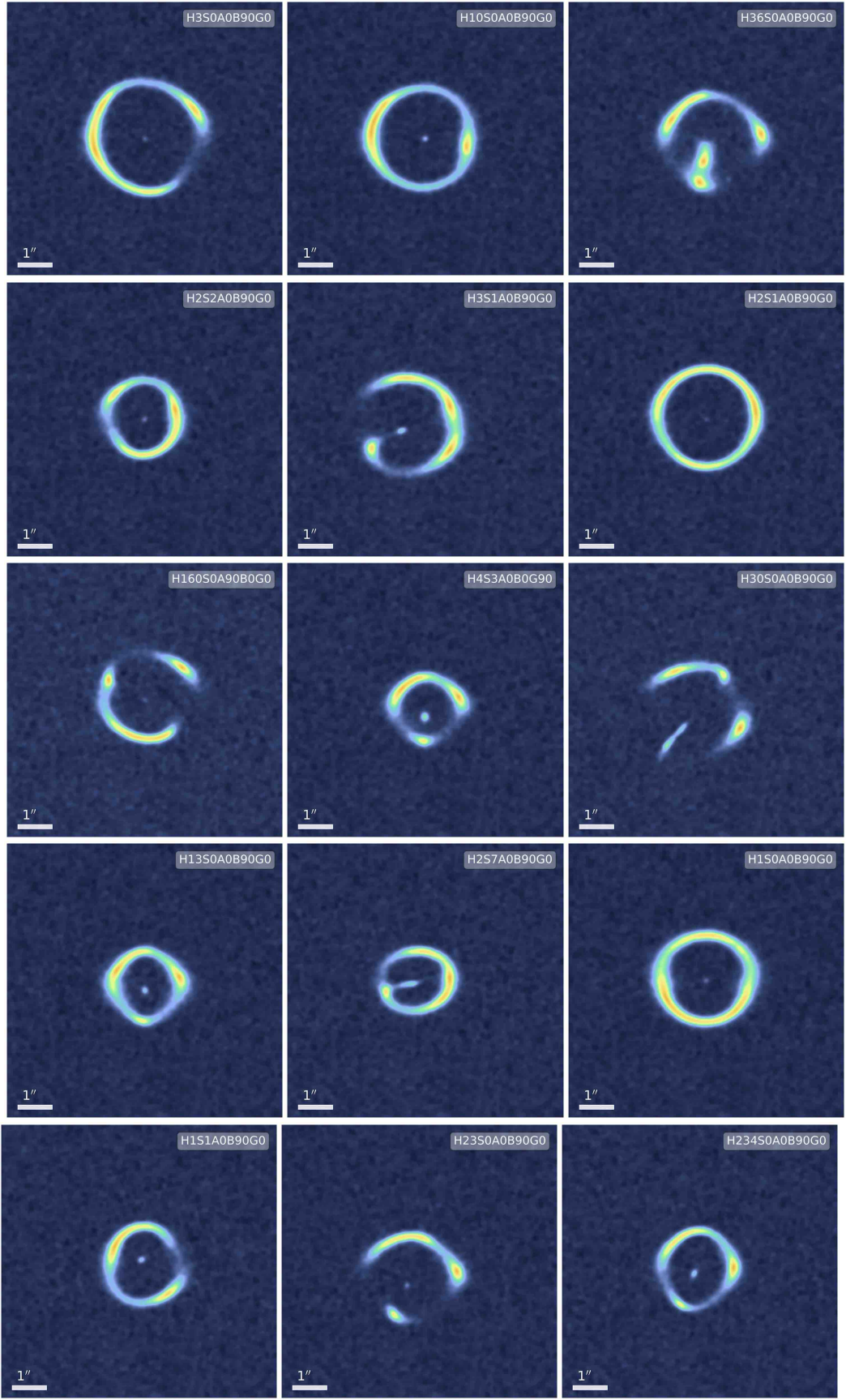

  \SEAGLEmosaic{data}{height=\pheight}
  \caption{ The SEAGLE-simulated lens data.  The pictures show the
    lensed images without the lens in arbitrary units of brightness.
    They were the only information provided in the blind study.  The
    scale bar on the lower left in each panel shows the physical scale
    in arcseconds.  }\label{fig:data}
\end{figure*}

  \begin{figure*}
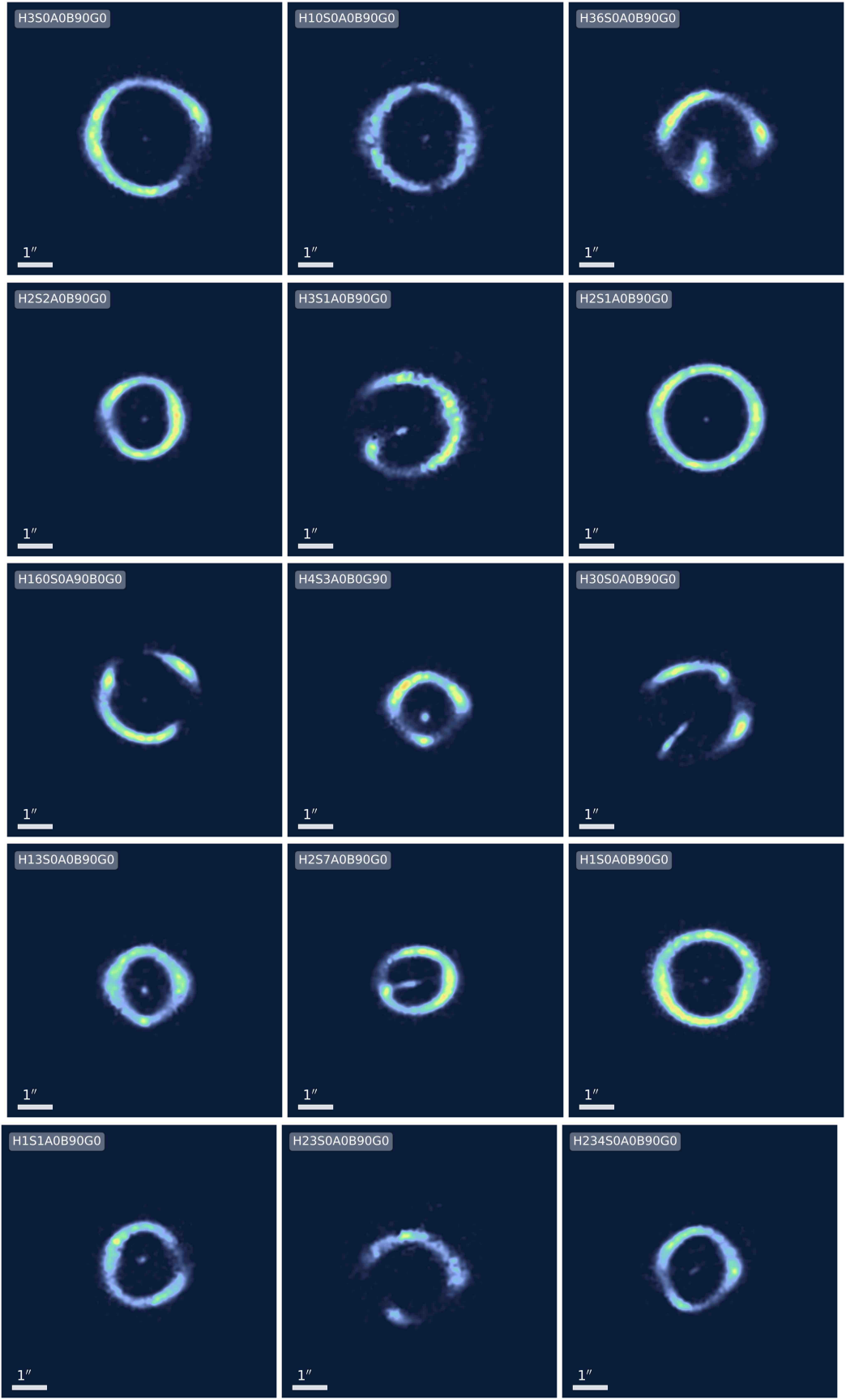

  \SEAGLEmosaic{synth}{height=\pheight}
  \caption{ Synthetic images of the 15 reconstructed lenses using the
    best-fitting model from the ensemble solution.  Scales are
    identical to the corresponding pictures in Figure~\ref{fig:data}.
  }\label{fig:synth}
\end{figure*}

  \begin{figure*}
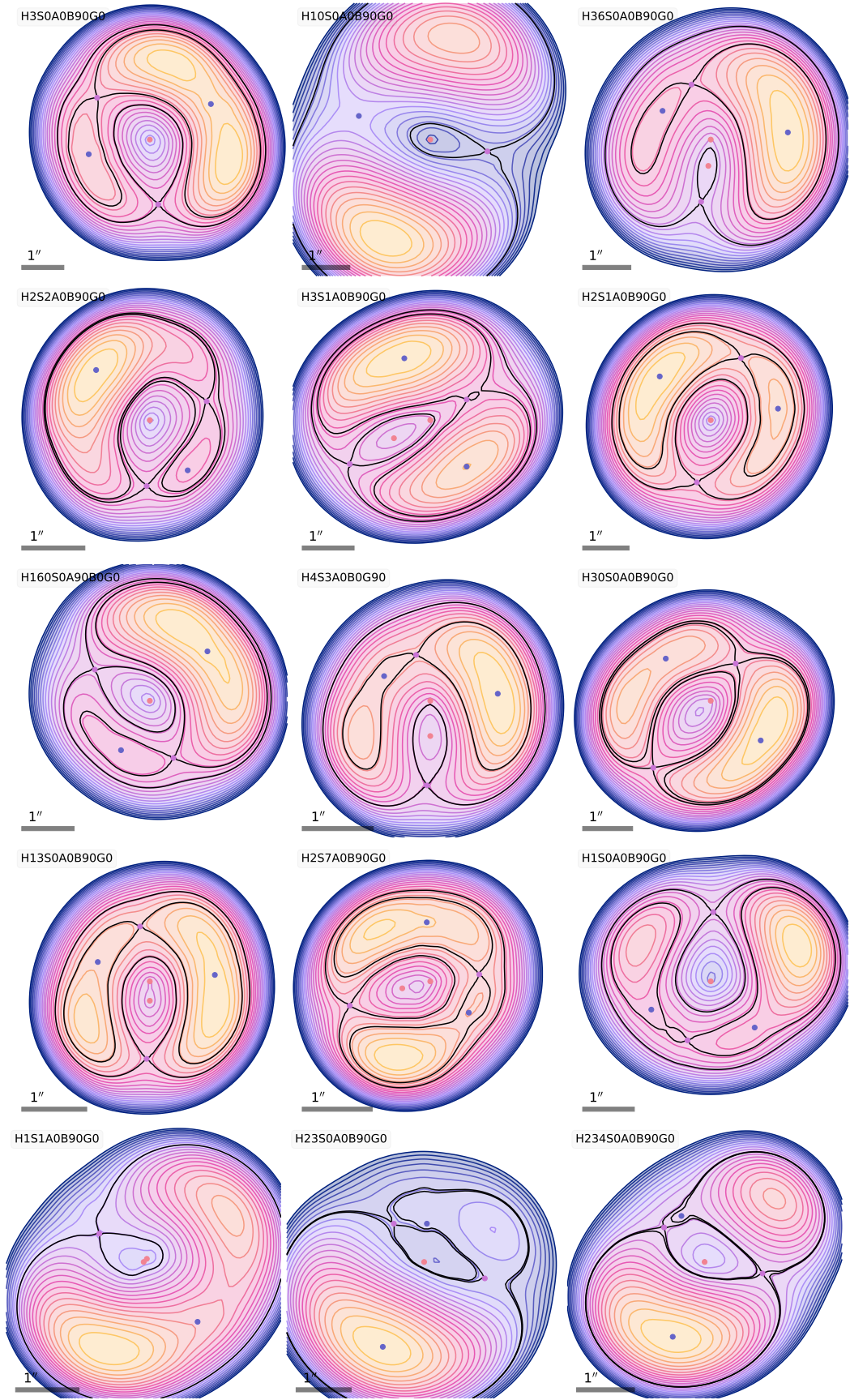

  \SEAGLEmosaic{arriv}{height=\pheight}
  \caption{Arrival-time surfaces of the ensemble average of the 15
    reconstructed lenses.  Contours passing through saddle points are
    in black. The lens centre is indicated by a red dot, while the
    image-position constraints with minimum, saddle, and maximum
    parity are indicated by blue, purple, and red dots
    respectively. It is easy to identify unrealistic models by looking
    for irregularities in the arrival-time surface
    contours. }\label{fig:arriv}
\end{figure*}

  \begin{figure*}
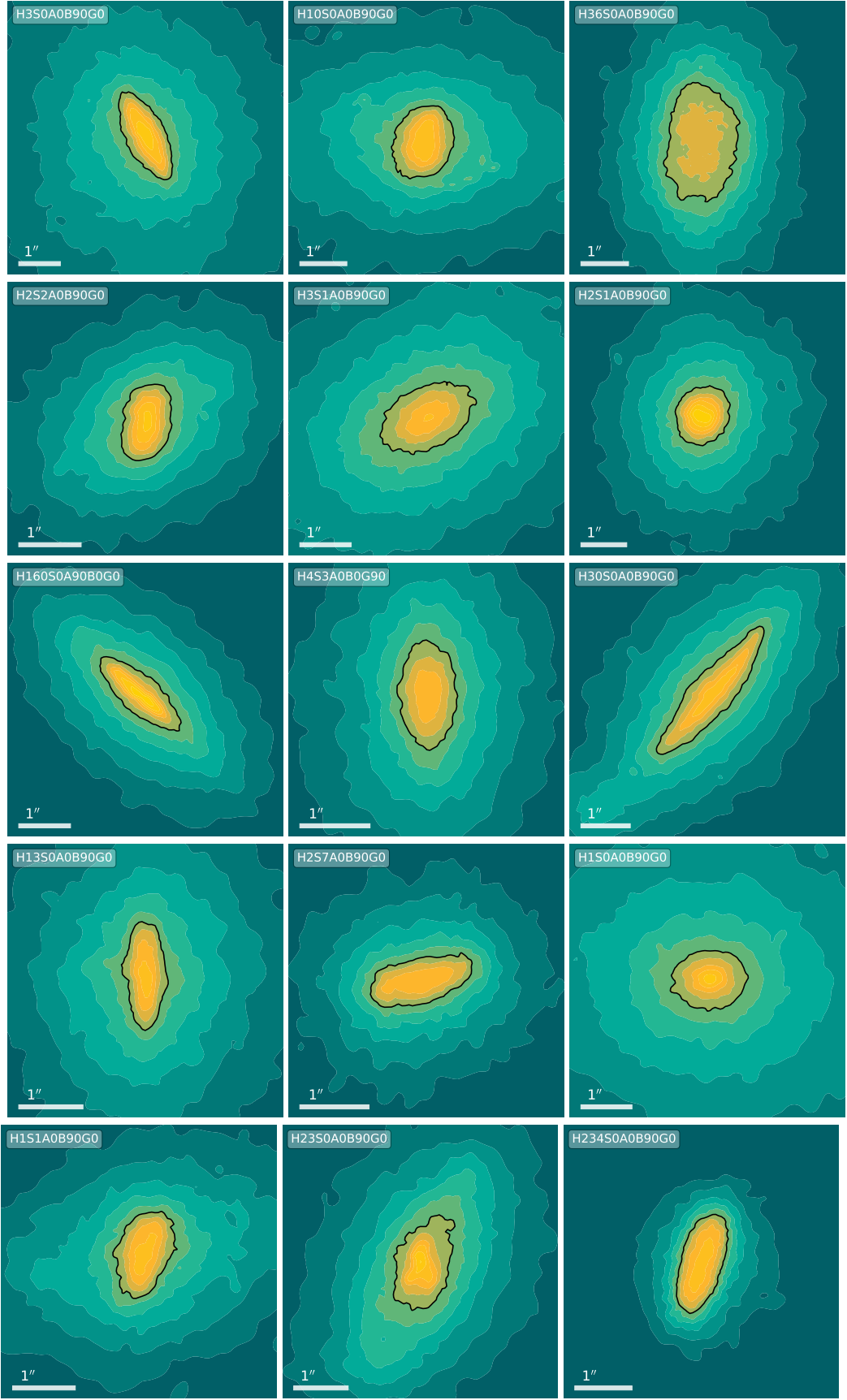

  \SEAGLEmosaic{kappa_true}{height=\pheight}
  \caption{The actual convergence maps of SEAGLE-simulated lenses.
    Black contours indicate a $\kappa = 1$. They were unblinded after the
    lens reconstructions were completed. }\label{fig:kappa_true}
\end{figure*}

  \begin{figure*}
  \SEAGLEmosaic{kappa}{height=\pheight}
  \caption{Model convergence maps (ensemble-averages) of all the
    lenses.  Black contours indicate a $\kappa = 1$.  Scales are identical
    to the corresponding pictures in Figure~\ref{fig:kappa_true}.
    Direct comparison of convergence maps usually fails, because they
    are affected by degeneracies. Nevertheless, the orientation of the
    lenses roughly match their SEAGLE counterparts. }\label{fig:kappa}
\end{figure*}

  \begin{figure*}
  \includegraphics[height=0.93\textheight]{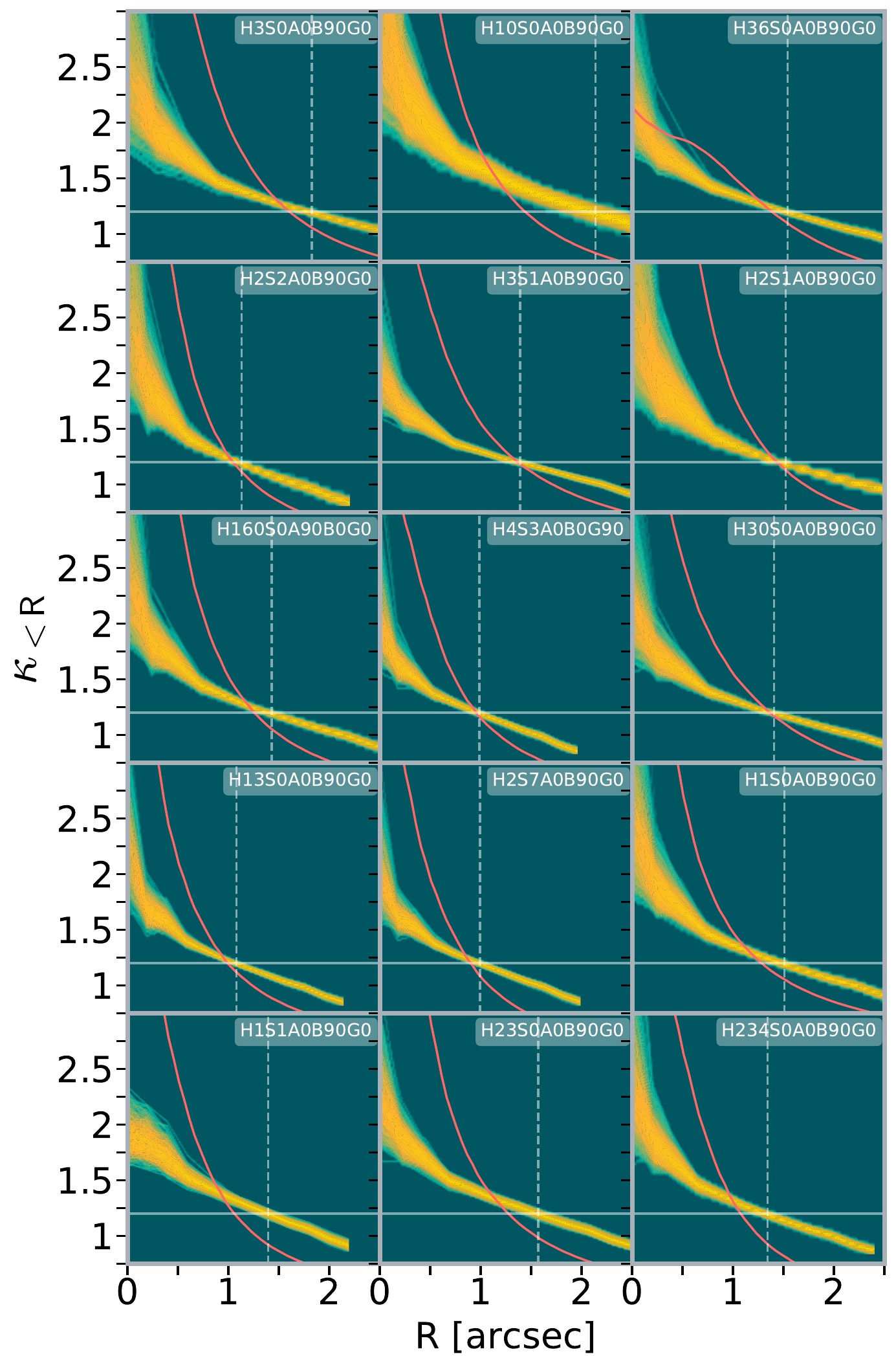}
  \caption{ Circularly averaged enclosed surface density profiles of all
    reconstructed lenses.  In other words, the plots show the enclosed mass as a
    mean $\kappa_{<\mathrm{R}}$ within a given projected radius from the lensing
    galaxies' center of mass.  The ensemble is represented with a coloured
    region with a gradient from green to yellow indicating its number density.
    The vertical line shows the approximate Einstein radius of the ensemble
    average.  The red line shows the same profile for the actual convergence
    map.  The Einstein radii of quads are in good agreement with the SEAGLE
    lenses.  It is harder to find the correct Einstein radius for doubles.
    }\label{fig:kappa_profile}
\end{figure*}

  \begin{figure*}
  \includegraphics[height=0.95\textheight]{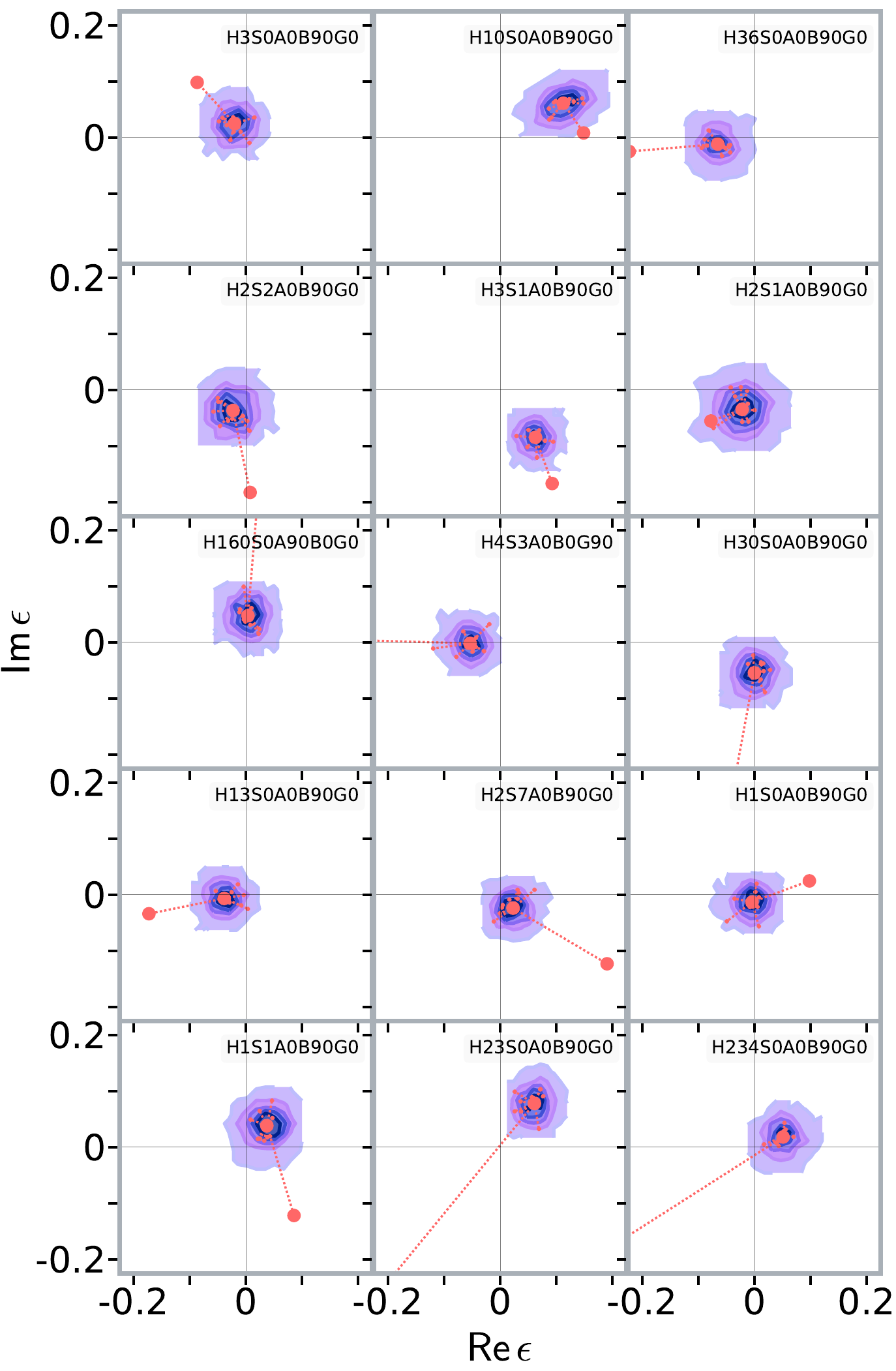}
  \caption{ Complex ellipticity model distribution within each
    ensemble model, compared to the actual complex ellipticity of the
    SEAGLE-simulated lenses.  The ellipticity of the SEAGLE lens is
    indicated with a large red dot which is connected to a graph with
    a center node indicated with another red dot representing the
    ellipticity of the model's ensemble average. The graphs leaf nodes
    are 20 randomly sampled points from the ensemble.  In most cases,
    the models are too round, however their position angles roughly
    match.  }\label{fig:ellipticity}
\end{figure*}

  \begin{figure*}
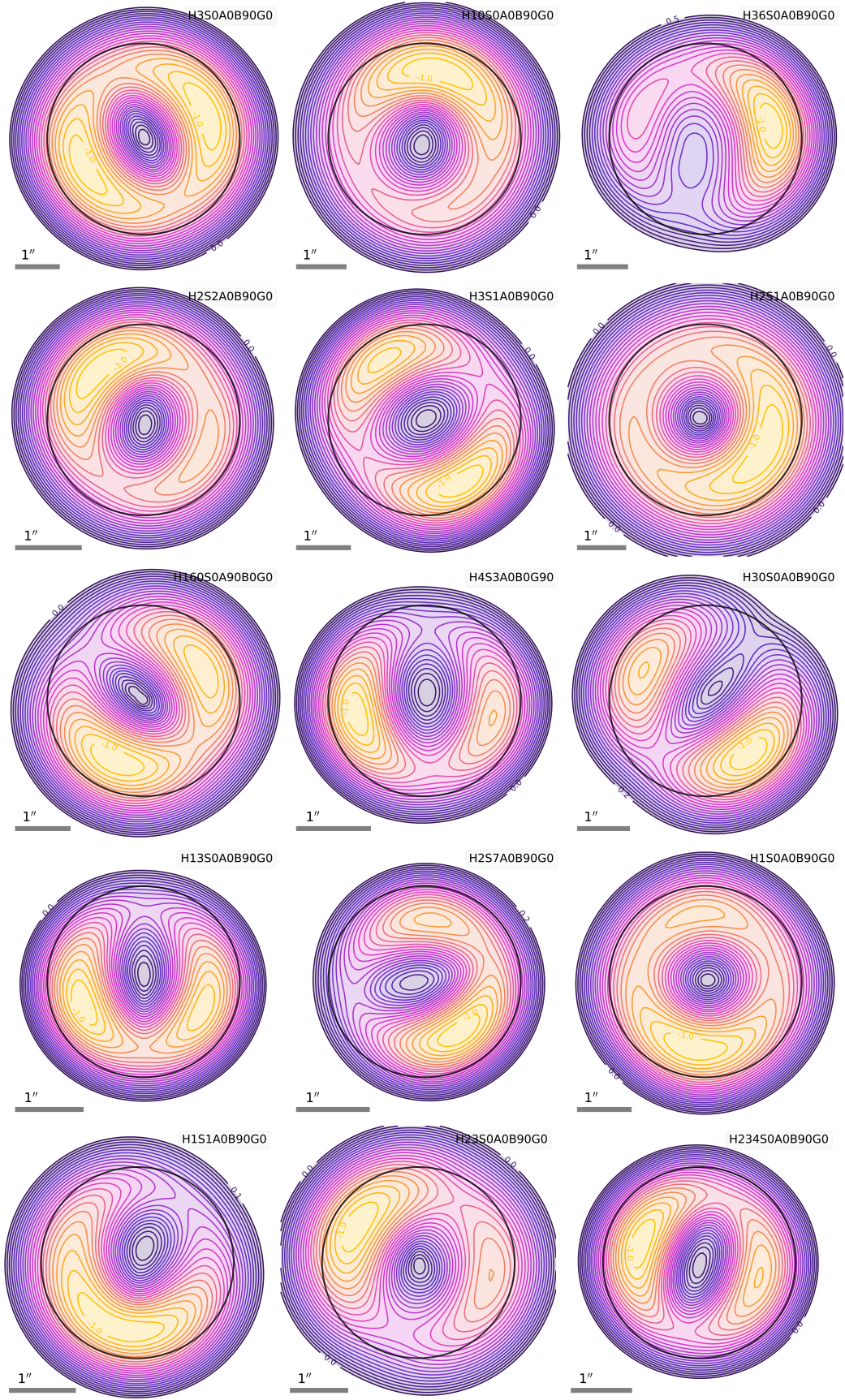

  \SEAGLEmosaic{roche_potential_true}{height=\pheight}
  \caption{ The actual Roche potentials of the SEAGLE-simulated
    lenses.  The black circle indicates the area within which the
    scalar product~\eqref{eq:Roche_scalar} was evaluated.  The
    potentials were shifted and scaled such that the center has a
    value of 0, and the global minimum a value of -1.  This way, they
    offer a convenient way of comparing different models with the
    effect of the steepness degeneracy taken out.
  }\label{fig:roche_true}
\end{figure*}

  \begin{figure*}
  \SEAGLEmosaic{roche_potential_ens_avg}{height=\pheight}
  \caption{ The Roche potentials of the 15 reconstructed lenses using
    the ensemble averaged model.  All the scales are identical to the
    corresponding pictures in
    Figure~\ref{fig:roche_true}. }\label{fig:roche}
\end{figure*}

  \begin{figure*}
  \includegraphics[height=0.95\textheight]{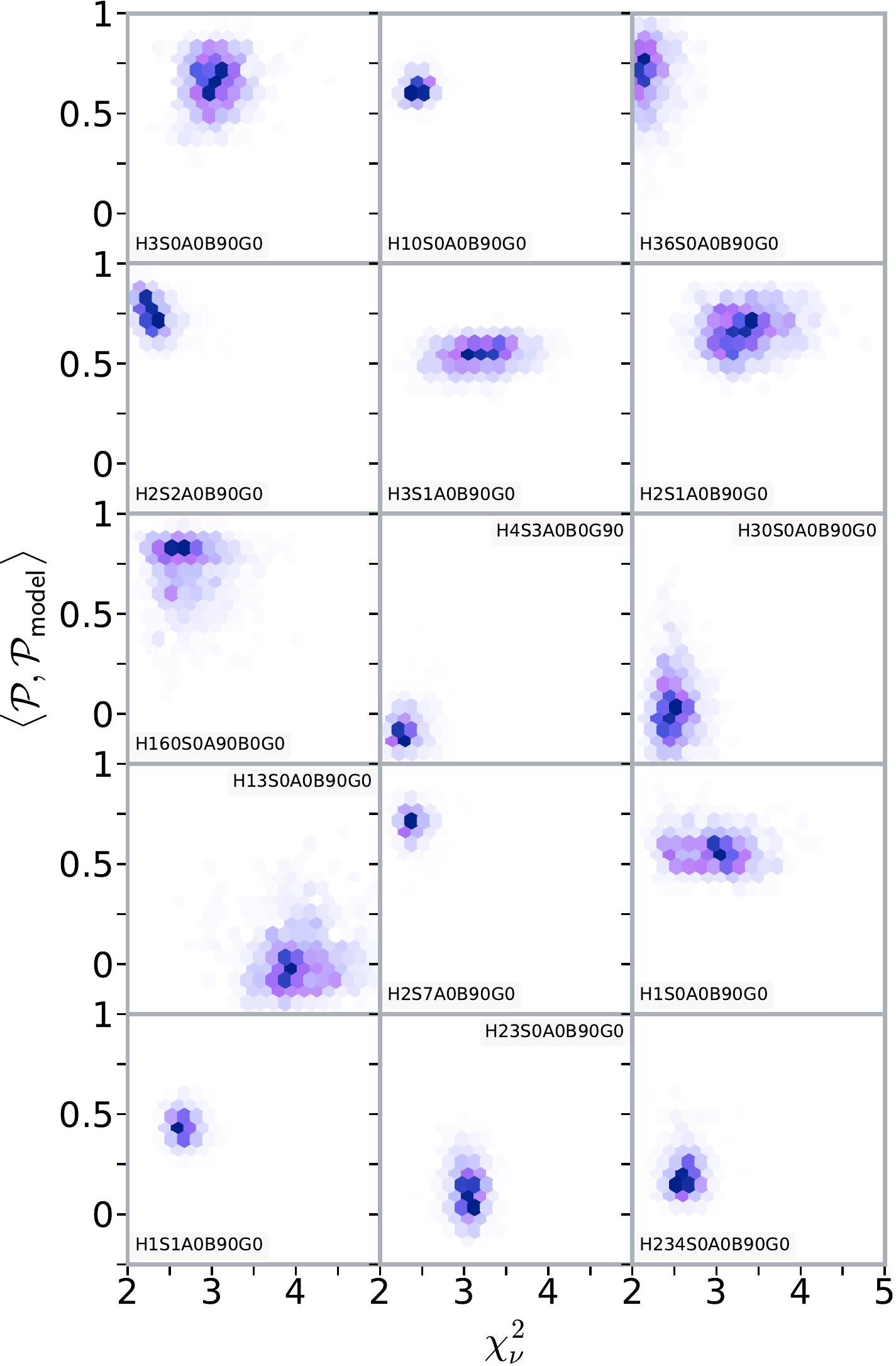}
  \caption{ Hex-binned plots of distributions of reduced $\chi^{2}_{\nu}$ from
    the lens image reconstructions againts the distribution of the scalar
    product of Roche potentials for each ensemble model, see
    equations~\eqref{eq:chi2} and~\eqref{eq:Roche_scalar}.  Particularly good
    lens recovery is indicated by a scalar product close to 1 and a low
    $\chi^{2}$.}\label{fig:chi2_VS_roche_scalar}
\end{figure*}

  \begin{figure}
  \includegraphics[width=0.37\textwidth]{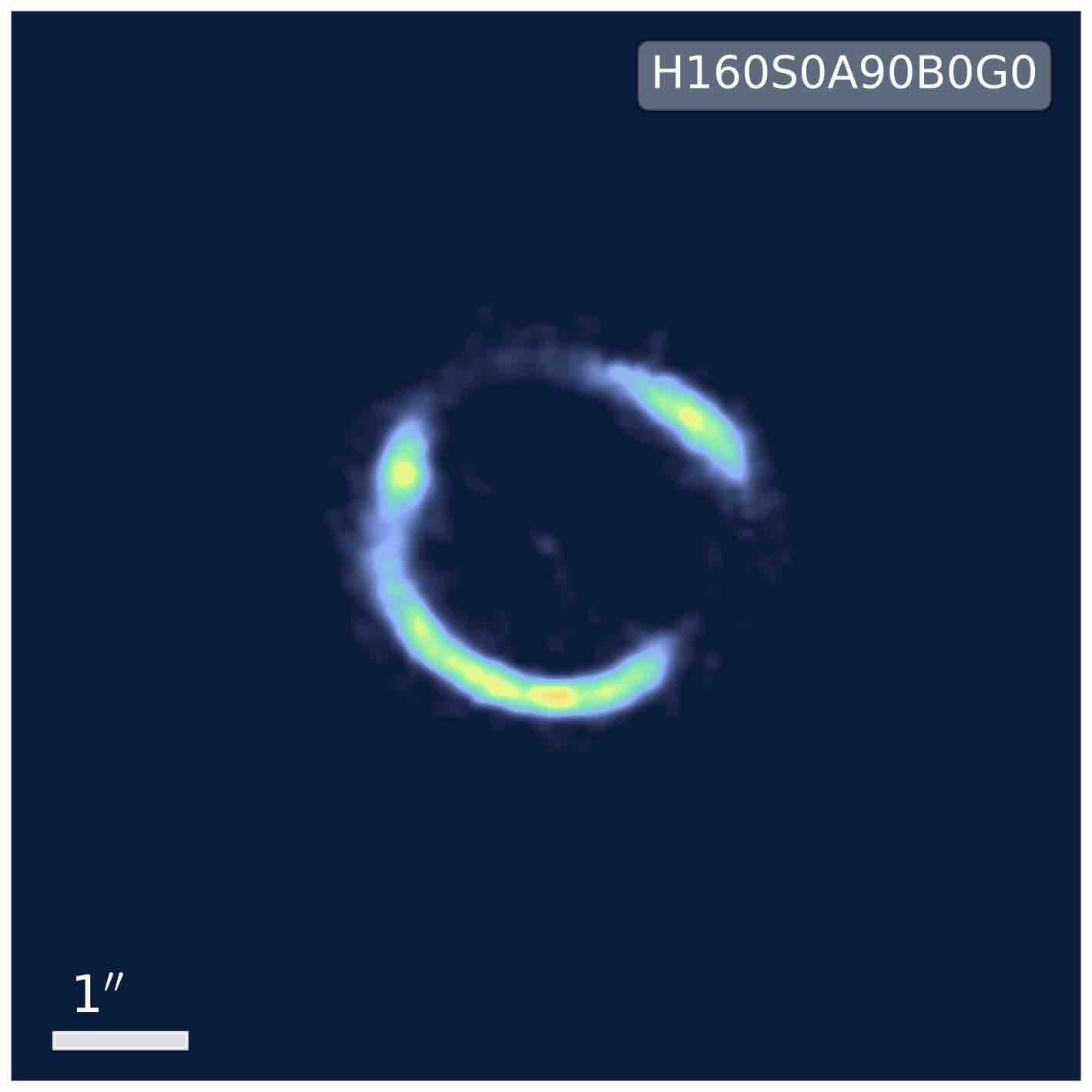}
  \includegraphics[width=0.37\textwidth]{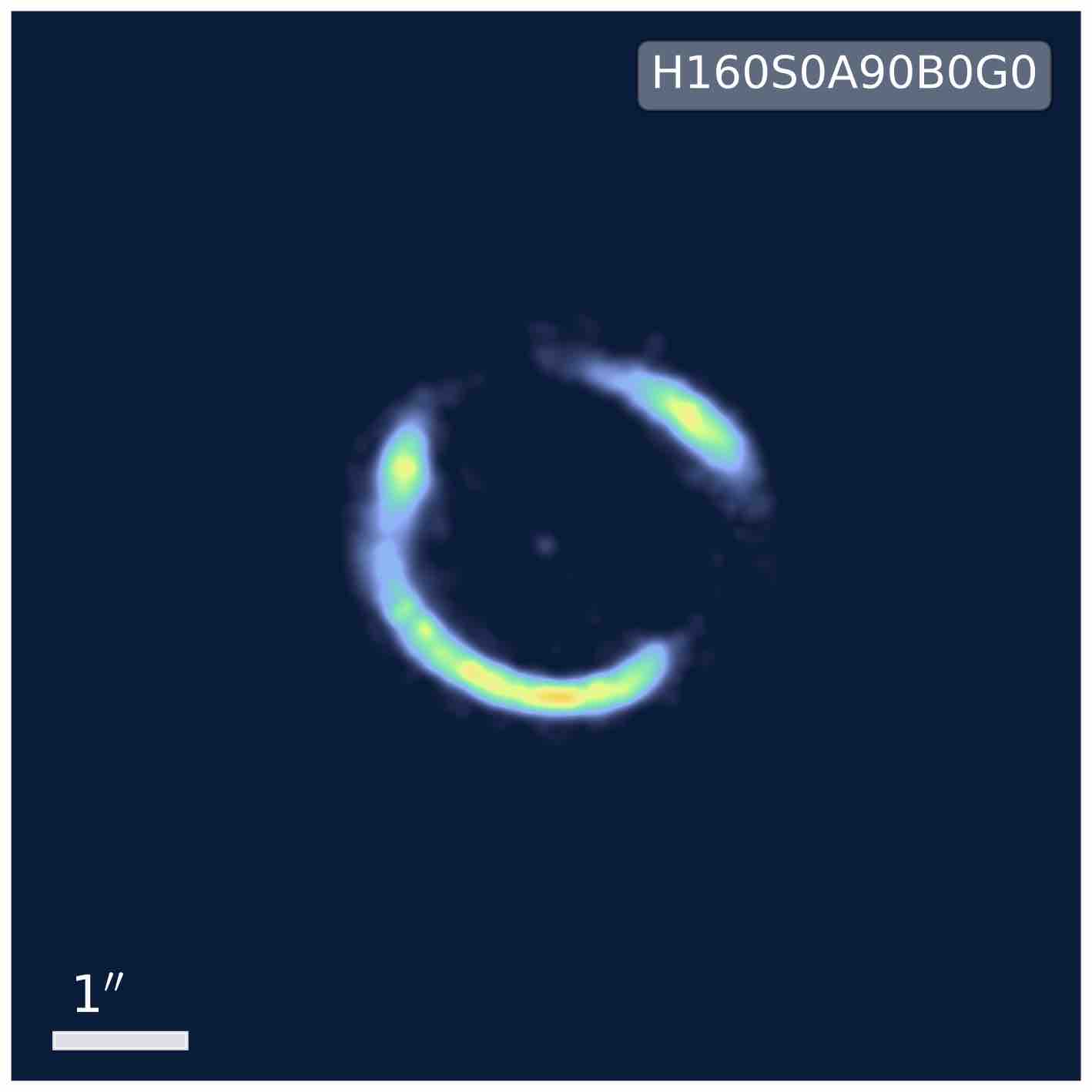}
  \includegraphics[width=0.37\textwidth]{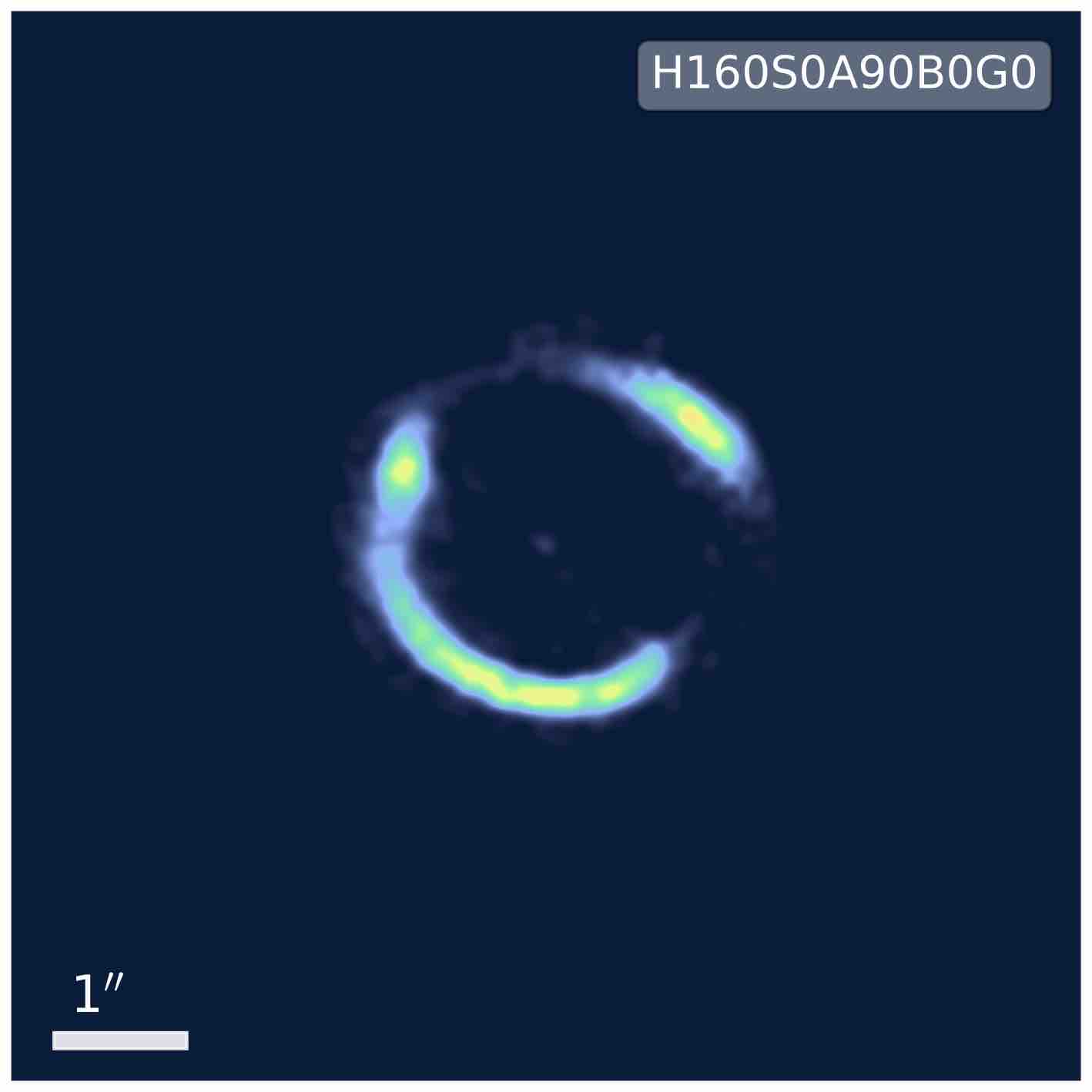}
  \caption{ Alternative synthetic images of the lens \protect\lens{13}\
    (cf.~middle-left panel in Figure~\ref{fig:synth}).  Here, the image
    on top was generated from the unfiltered ensemble consisting of
    1000 models.  The middle shows the image generated with an
    ensemble containing 100 models with lowest $\chi^{2}$, and the
    bottom image shows the synthetic image from an ensemble containing
    79 models which is the intersection of the set of lowest
    $\chi^{2}$ and the scalar products of the Roche potentials closest
    to 1.  The middle image shows a visible, albeit small, improvement
    compared to the top image, whereas the bottom image only shows
    slightly better relative brightness between the lens images
    compared to the middle, but no discernible improvement otherwise.
  }\label{fig:filtered}
\end{figure}

\end{section}

\label{lastpage}
\clearpage

\end{document}